\documentclass[10pt,journal,compsoc]{IEEEtran}

\IEEEoverridecommandlockouts
\ifCLASSOPTIONcompsoc
    \usepackage[nocompress]{cite}
\else
    \usepackage{cite}
\fi
\ifCLASSOPTIONcompsoc
    \usepackage[caption=false,font=footnotesize,labelfont=sf,textfont=sf]{subfig}
\else
    \usepackage[caption=false,font=footnotesize]{subfig}
\fi
\ifCLASSOPTIONcaptionsoff
    \usepackage[nomarkers]{endfloat}
\let\MYoriglatexcaption\caption
\renewcommand{\caption}[2][\relax]{\MYoriglatexcaption[#2]{#2}}
\fi
\usepackage[T1]{fontenc}
\usepackage{url}
\usepackage{subfig}
\usepackage{color,xcolor}
\usepackage{dsfont}
\usepackage{bm}
\usepackage[pdftex]{graphicx}
\usepackage{array,amsmath,amssymb,amsfonts}
\usepackage{amsthm}
\usepackage{subeqnarray}
\usepackage{cases}
\usepackage{booktabs}
\usepackage[ruled,lined,linesnumbered,commentsnumbered,longend]{algorithm2e}
\makeatletter
\newcommand{\removelatexerror}{\let\@latex@error\@gobble}
\makeatother

\SetCommentSty{mycommfont}
\renewcommand{\vec}[1]{\boldsymbol{#1}}

\DeclareMathOperator*{\argmax}{argmax}

\newtheorem{theorem}{\textbf{Theorem}}

\newtheorem{proposition}{\textbf{Proposition}}
\newtheorem{definition}{\textbf{Definition}}
\def\BibTeX{{\rm B\kern-.05em{\sc i\kern-.025em b}\kern-.08em T\kern-.1667em\lower.7ex\hbox{E}\kern-.125emX}}
\vspace{-0.3cm}
\IEEEoverridecommandlockouts

\begin{document}

\title{DPoS: Decentralized, Privacy-Preserving, and Low-Complexity Online Slicing for Multi-Tenant Networks}
\author{Hailiang~Zhao,
        Shuiguang~Deng,~\IEEEmembership{Senior~Member,~IEEE,}
        Zijie~Liu,
        Zhengzhe~Xiang,~\IEEEmembership{Member,~IEEE,}
        Jianwei~Yin,
        Schahram~Dustdar,~\IEEEmembership{Fellow,~IEEE},
        and~Albert~Y.~Zomaya,~\IEEEmembership{Fellow,~IEEE}%
\IEEEcompsocitemizethanks{
  \IEEEcompsocthanksitem H. Zhao, S. Deng, Z. Liu, and J. Yin are with 
  the College of Computer Science and Technology, Zhejiang University, Hangzhou 310058, China.\\
  E-mail: \{hliangzhao, dengsg, liuzijie, zjuyjw\}@zju.edu.cn.
  \IEEEcompsocthanksitem Z. Xiang is with Zhejiang University City College, Hangzhou 310015, China.
  E-mail: xiangzhengzhe@zju.edu.cn.
  \IEEEcompsocthanksitem S. Dustdar is with the Distributed Systems Group, TU Wien, 1040 Vienna, Austria.
  E-mail: dustdar@dsg.tuwien.ac.at.
  \IEEEcompsocthanksitem A. Y. Zomaya is with the School of Computer Science, University of Sydney, Sydney, 
  NSW 2006, Australia.\\ 
  E-mail: albert.zomaya@sydney.edu.au.
  \IEEEcompsocthanksitem Shuiguang Deng is the corresponding author.}%
}

\IEEEtitleabstractindextext{%
\begin{abstract}
    Network slicing is the key to enable virtualized resource sharing among vertical industries in the era of 5G communication. 
    Efficient resource allocation is of vital importance to realize network slicing in real-world business scenarios. To deal 
    with the high algorithm complexity, privacy leakage, and unrealistic offline setting of current network slicing algorithms, 
    in this paper we propose a fully decentralized and low-complexity online algorithm, DPoS, for multi-resource slicing. We first 
    formulate the problem as a global social welfare maximization problem. Next, we design the online algorithm DPoS based on 
    the primal-dual approach and posted price mechanism. In DPoS, each tenant is incentivized to make its own decision based on 
    its true preferences without disclosing any private information to the mobile virtual network operator and other tenants. 
    We provide a rigorous theoretical analysis to show that DPoS has the optimal competitive ratio when the cost function of each 
    resource is linear. Extensive simulation experiments are conducted to evaluate the performance of DPoS. The 
    results show that DPoS can not only achieve close-to-offline-optimal performance, but also have low algorithmic overheads. 
\end{abstract}

}

\maketitle

\IEEEdisplaynontitleabstractindextext

\IEEEpeerreviewmaketitle

\IEEEraisesectionheading{\section{Introduction}\label{sec1}}

\IEEEPARstart{5}{G} creates tremendous opportunities for social digitalization and industrial interconnection. On top of the physical 
infrastructure, diversified service requirements (eMBB, mMTC, and uRLLC) can be met in the service-oriented end-to-end network slicing 
(E2E-NS) architecture. The E2E-NS architecture supports both the co-existent accesses of multiple standards (5G, LTE, and Wi-Fi), and 
the coordination between different site types (macro cell, micro cell, and pico cell base stations), which is mainly attributed to the 
flexible orchestration and on-demand deployment of virtualized network functions (VNFs) \cite{5G-archi}\cite{ns-survey2}\cite{ns-survey3}. 

\begin{figure}[htbp]
    \centerline{\includegraphics[width=1.7in]{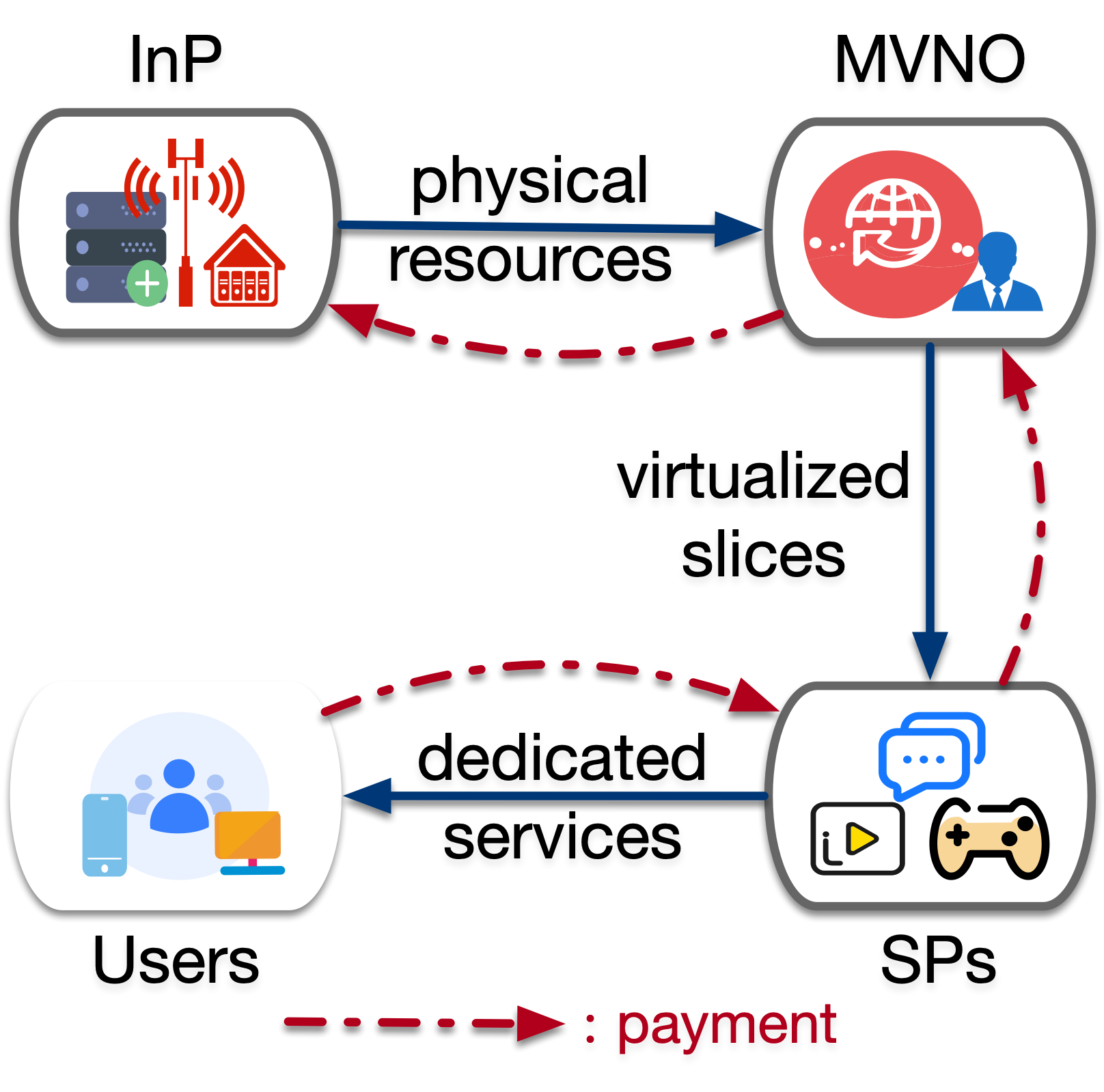}}
    \caption{Business players involved in network slicing and how does the process works.}
    \label{fig1}
\end{figure}

The substantive characteristics of the E2E-NS architecture is \textit{cloudification}. It involves the transformation from traditional 
hardbox network functions to the all-on-cloud management \& control planes \cite{ns-survey}. In this architecture, network slicing is 
the key to enabling networking capabilities for vertical industries. Many business players, such as infrastructure providers 
(InPs), mobile network operators (MNOs), cloud providers (one kind of InPs actually), edge \& cloud service providers (a.k.a. \textit{tenants}), 
service subscribers (i.e. \textit{users}), service brokers and mobile virtual network operators (MVNOs) are involved\cite{ns-business}
\cite{mobicom-slicing-paper}. For the scenario considered in this paper, the InP offers the physical network infrastructure to the MVNO by 
leasing or selling and is responsible for hardware upgrades and maintenance. After having control of the physical networks, the MVNO 
virtualizes the network resources, divides each kind of resource into slices, and rents them to the tenants according to tenants' demand. 
Therewith, each tenant creates service instances based on its slices, and provides services to its subscribers. Normally, the level of 
services are stated in Service Level Agreements (SLAs). SLAs define the metrics to measure and show if the expected \textit{quality of service} 
(QoS) is achieved or not. The process\footnote{To be clear, this is not the only business model in network slicing. In some scenarios, 
the MNOs are the owners and maintainers of physical resources. They create slices on top of the resources, which will be offered to the MVNOs 
to perform services to subscribers. For all this, the network slicing model provided in this paper is applicative.} is illustrated in 
Fig. \ref{fig1}.

The key problem underlying network slicing is \textit{efficient resource allocation} for VNFs \cite{nfv}\cite{nfv-offline}, 
which is algorithmicaly NP-hard \cite{ns-algorithm-pers}. There has been lots of research done so far for different scenarios, including 
slicing the radio access networks (RANs) \cite{ran-slice}\cite{edge-slice2}\cite{tsc-edge}\cite{tmc-edge}, the 
core networks (5GCs) \cite{core-slice}\cite{core-add1}\cite{core-add3}, and the federated edge \cite{edge-slice} \cite{resource-manage-DQN} \cite{tpds-burst}, 
etc. In these cases, survivability constraints, heterogeneous QoS requirements, geographical limitations, and other \textit{scenario-specific} 
constraints are taken into considerations to formulate complicated combinatorial non-convex problems. To solve them, the most typical 
and general class of works are based on fine-tuned heuristics \cite{heuristic} or deep machine learning models such as deep $Q$-network (DQN) 
\cite{resource-manage-DQN}\cite{DL}. These algorithms can achieve (approximately) optimal solutions and make the communication systems smart 
and intelligent \cite{add3}\cite{edge-intelligence}. However, they are usually complex and do not scale with the types of resources and the 
number of tenants. Take Deep $Q$-Network (DQN) as an example, it could take days even weeks for obtaining \textit{not-particularly-good} actions 
even though the state and action spaces have been discretized. Although several reinforcement learning methods can avoid 
privacy leakage, such as \cite{add1} and \cite{add2}, the centralized algorithms are generally built on the complete knowledge regarding 
all preferences of involved business players, including the monetary budget of tenants, the number and purchasing-power of service subscribers, 
etc. The formulation of the centralized optimization problem itself is a detriment on privacy and trade secrets.

To avoid insufferable complexity and privacy leakage, in recent years, many researchers establish network slicing models based 
on standard economic frameworks, such as \textit{Fisher markets} \cite{fisher1}\cite{fisher2}, and different auction-based mechanisms, 
such as the VCG-Kelly mechanisms \cite{VCG}\cite{vcg-kelly}. In these works, all tenants get together and bid for maximizing their 
profits. For instance, Wang et al. studied the relationship between resource efficiency and profit maximization and developed an 
optimization framework to maximize net social welfare \cite{s1}. Similarly, Jiang et al. addressed a joint resource and revenue 
optimization problem and solved it with the auction mechanisms \cite{s3}. Furthermore, some works resort to game theory to 
model tenants' and MVNOs' strategic (or non-strategic) behaviors, and take the \textit{price of anarchy} (PoA) to analyze the efficiency 
of potentially existent Nash equilibrium (NE) \cite{zhang2011game}. For instance, Caballero et al. studied the resource allocation 
mechanism by formulating a network slicing game \cite{congest-game-slice}. They proved that when the game associated with strategic 
behavior of tenants, i.e., adjusting their preferences depending on perceived resource contention, convergence to a Nash equilibrium 
(\textit{under some specific conditions}) can be achieved. Luu et al. also study a network slicing game, but under specific constraints 
of RAN \cite{ran-slice}. Generally, auction mechanisms are efficient and scalable to diversified service requirements. However, most of 
these auction-based works are designed under an \textit{offline} setting, i.e., the MVNO knows the willingness to bid and many other 
private information of all tenants during each bidding round. Besides, a tenant's partial private information might be disclosed to 
all the remaining tenants. Nevertheless, this may not possible in many real-world business transactions because it is rare that all 
the tenants negotiate the rental business details simultaneously. The MVNO should not know anything about the arrival 
sequence of tenants, much less the private information of the served users of each tenant. It should only have the knowledge on the 
resource surplus and the attributes saved in the generic network slice templates (GSTs) \cite{nst}. In addition, a tenant private 
information should not be available to the other tenants.

The above analysis shows that auction mechanisms may not be ideal for online network slicing problems. In addition to the above 
reasons, auctions take time and require multiple communication rounds between the MVNO and the tenants \cite{auction-or-posted}. 
They may have poor performances when the distribution of bidders' arrival instance is unknown \cite{posted-mechanism}. By contrast, 
take-it-or-leave-it, i.e., \textit{posted price}, is a more practical option for online settings. Therefore, in this paper, we 
design an online slicing algorithm based on the posted price mechanism. A decentralized, low-complexity, and privacy-preserving 
algorithm, \textsf{DPoS}, mainly based on previous theoretical works on the online primal-dual algorithms \cite{primal-dual2}, 
\cite{online-mechanism}, and \cite{primal-dual1}, is proposed. Specifically, we extend the basic model proposed in \cite{online-mechanism} 
into multi-resource scenarios. \textsf{DPoS} is consists of two parts, \textsf{DPoS-MVNO} (agent for the MVNO) and \textsf{DPoS-TNT}$_n$ 
(agent for the $n$-th tenant), with a complexity of $O(N C)$ and $O(C)$, respectively. Here $N$ is the number of tenants, and $C$ is 
the number of type of resources. \textsf{DPoS} runs in a fully decentralized way. Each time a new tenant $n$ arrives, \textsf{DPoS-TNT}$_n$ 
decides to rent the demand resources or not according to the rental prices of each type of resource, published by \textsf{DPoS-MVNO} 
beforehand. Therewith, \textsf{DPoS-TNT}$_n$ sends the decision and payment (if tenant $n$ has the willingness to pay) to \textsf{DPoS-MVNO}. 
Then, \textsf{DPoS-MVNO} checks whether the resource surplus can satisfy tenant $n$ and inform \textsf{DPoS-TNT}$_n$ the transaction 
is succeeded or failed. Note that each tenant may experience different prices on the same kind of resource, which depends on the 
pricing mechanism the MVNO adopts. In the above procedure, only a small flow of privacy-irrelevant information are transferred between 
the MVNO and each tenant. No information are transferred among tenants. Trade secrets, especially the information of service subscribers 
and the pricing policies of tenants, will not be disclosed.

Our main contributions are summarized as follows.
\begin{itemize}
    \item We design a decentralized, privacy-preserving online network slicing algorithm, \textsf{DPoS}. This algorithm 
    enjoys low complexity, and it is practicable under diversified multi-resource requirements. Trade secrets and related private 
    information can be fully preserved.
    \item We find that, when the cost function of each resource is linear, \textsf{DPoS} achieves the \textit{optimal} competitive 
    ratio over all the online algorithms for the maximization of social welfare.
    \item We verify the superiority of \textsf{DPoS} from multiple angles, including social welfare achieved, 
    cross-agent communication data size, algorithm execution time, etc. The experimental results show that \textsf{DPoS} not only achieves 
    close-to-offline-optimal performance, but also has low algorithmic overheads. 
\end{itemize}

The remainder of the paper is organized as follows. Sec. \ref{sec2} presents the system model and formulates the global offline problem. 
Sec. \ref{sec3} demonstrates the design details of the algorithm \textsf{DPoS}. Theoretical analysis on the competitive ratio is provided 
in Sec. \ref{sec4}. The experiment results are demonstrated in Sec. \ref{sec5}. Sec. \ref{sec6} reviews related works and Sec. \ref{sec7} 
concludes this paper. 

\section{Problem Formulation}\label{sec2}
To simplify the notations without damaging its economic structure, our model concerns one InP, one MVNO, several tenants and each tenant's 
served users. Our model and algorithm can be directly adapted to multi-MVNO multi-InP scenarios. We consider the scenario where 
multiple network slices are built upon an SDN/NFV-enabled 5G network infrastructure, which is rented from the InP by the MVNO. 
Roughly, physical resources in this infrastructure can be divided into computation, storage, and forwarding/bandwidth. At great length, physical 
resources are usually organized as a \textit{weighted directed graph} \cite{s1} \cite{resource-allocation-infocom17}, where the \textit{node} 
can be base station at the access, forwarding router in the bearer, and physical machine or virtual machine in the regional datacenters, and 
the \textit{edge} is a directed link with certain propagation speed. Each node is the carrier of VNFs with different capabilities 
while link has unique bandwidth and data transfer rate. To ensure the universality of the model, we do not add any specific limitation and simply 
use $\mathcal{C} \triangleq \{1, ..., C\}$ to denote the set of resources. Without loss of generality, the capacity limit of each resource is 
normalized to be $1$.

\begin{table}[htbp]   
    \begin{center}
        \caption{\label{key}Summary of key notations.}   
    \begin{tabular}{l|l}    
        \toprule
        {\textsf{\textbf{Notation}}}& {\textsf{\textbf{Description}}}\\[+0.1mm]
        \midrule
        $\mathcal{C}$ & The set of network resources\\[+0.7mm]
        $\mathcal{N}$ & The set of tenants\\[+0.7mm]
        $\mathcal{S}_n$ & The set of users of tenant $n \in \mathcal{N}$\\[+0.7mm]
        $\{d_n^c\}_{\forall c \in \mathcal{C}}$ & Resource demands of tenant $n$\\[+0.7mm]
        $v_n$ & The revenue in estimation of tenant $n$\\[+0.7mm]
        $e_n^c$ & The valuation density of tenant $n$\\[+0.7mm]
        $\underline{p_c} \textrm{ and } \overline{p_c}$ & The lower (upper) bound of earning density\\[+0.7mm]
        $x_n \in \{0, 1\}$ & The decision variable of tenant $n$\\[+0.7mm]
        $\pi_n$ & The payment made by tenant $n$\\[+0.7mm]
        $y_c \in [0, 1]$ & The resource rent out of type $c$\\[+0.7mm]
        $\{f_c\}_{\forall c \in \mathcal{C}}$ & Non-decreasing zero-startup cost functions\\[+0.7mm]
        $\underline{\varsigma_c}$ and $\overline{\varsigma_c}$ & The derivative of $f_c(\cdot)$ at point $0$ and $1$\\[+0.7mm]
        $\{\tilde{f}_c\}_{\forall c \in \mathcal{C}}$ & The extended cost functions\\[+0.7mm]
        $\{F_{p_c}\}_{\forall c \in \mathcal{C}}$ & The profit functions\\[+0.7mm]
        $\{h_{c}\}_{\forall c \in \mathcal{C}}$ & The maximum profit functions\\[+0.7mm]
        $\psi_n \textrm{ and } p_c$ & The dual variables corresponding to $x_n$ and $y_c$\\[+0.7mm]
        $\{\phi_c\}_{\forall c \in \mathcal{C}}$ & The pricing functions\\[+0.7mm]
        $\alpha$ & Competitive ratio of online algorithms\\[+0.7mm]
        \bottomrule   
    \end{tabular}  
    \end{center}
\end{table}

Let us use $\mathcal{N} \triangleq \{1, .., N\}$ to denote the set of tenants. In our model, each tenant requests one (and ony one) slice from 
the MVNO\footnote{In the following, we may interpret $n$ as the $n$-th tenant or the $n$-th slice. It depends on the content.}. Generally, a 
slice is a collection of different types of resources, the topology of which can be mapped onto the substrate network as a connected 
subgraph\footnote{There have been lots of works on the VNF placement and mapping \cite{nfv}\cite{core-add1}\cite{edge-slice}\cite{congest-game-slice}. 
But this is not the subject of this paper.}. We use $\{d_n^c\}_{\forall c \in \mathcal{C}}$ to denote the requirements of the $n$-th tenant (slice), 
where $d_n^c$ is the demand of resource of type $c \in \mathcal{C}_n \subseteq \mathcal{C}$, and
\begin{equation}
    d_n^c 
    \left\{
        \begin{array}{rl}
            > 0 & \textrm{if } c \in \mathcal{C}_n \\
            = 0 & \textrm{otherwise}.
        \end{array}
    \right.
\end{equation}

The traffic demand on the $n$-th slice is denoted by $\{f_s(\gamma, \tau)\}_{\forall s \in \mathcal{S}_n}$, where $s$ is a service subscriber 
from tenant $n$'s served users $\mathcal{S}_n$, and $f_s(\gamma, \tau)$ is a data flow with promissory data rate $\gamma$ and latency 
constraint $\tau$ from some source node to some destination node. A slice's consumption on resources is embodied in the execution of VNFs and 
the occupation of bandwidth. For tenant $n$, we define a function $\sigma_n: \{f_s(\gamma, \tau)\}_{\forall s \in \mathcal{S}_n} \to \mathbb{R}$ to calculate the 
payment of user $s$ for enjoying the data flow $f_s(\gamma, \tau)$. We regard $\sigma_n$ as \textit{private} because it involves business secrets 
of the tenant. The \textit{estimated} revenue of each tenant $n$ is from the payment of its service subscribers, which is defined as follows:
\begin{equation}
    v_n \triangleq \sum_{s \in \mathcal{S}_n} \varrho_s \cdot \sigma_n \Big(f_s(\gamma, \tau)\Big).
    \label{vn}
\end{equation}
In \eqref{vn}, $\varrho_s$ is the level of QoS for subscriber $s \in \mathcal{S}_n$, which is decided based on the commitment on delay torelant, reliability, 
isolation level, etc \cite{ns-game}. A full list of attributes can be found at \cite{nst}. Under normal circumstances, the higher the level of QoS, the 
faster data rate and tighter latency constraints on the data flow, which in turn leads to more resource consumption. 
Note that all the $\{d_n^c\}_{\forall c \in \mathcal{C}}$ are used to support the data flows $\{f_s(\gamma, \tau)\}_{\forall s \in \mathcal{S}_n}$. 
If we assume that the mapping function from data flow $f_s(\gamma, \tau)$ to the $c$-th resource occupation is 
$g_c: \{f_s(\gamma, \tau)\}_{\forall s \in \mathcal{S}_n} \to [0, d_n^c]$, we have the following identity:
\begin{equation*}
    d_n^c = \sum_{s \in \mathcal{S}_n} g_c \Big(f_s(\gamma, \tau)\Big), \forall c \in \mathcal{C}_n.
\end{equation*}

In practice, $v_n$ can be interpreted as the \textit{willingness-to-pay} of tenant $n$ for renting the required resources \cite{online-mechanism}. 
$\forall c \in \mathcal{C}, \forall n \in \mathcal{N}$, we define the \textit{earning density} $e_n^c$ as $v_n / d_n^c$. $e_n^c$ can be interpreted 
as the estimated revenue per unit of resource $c$ to the tenant $n$. Following \cite{online-mechanism}\cite{assumption1}, we define 
$\underline{p_c}$ and $\overline{p_c}$ as follows.
\begin{equation}
    \forall c \in \mathcal{C}:
    \left\{
        \begin{array}{l}
            \underline{p_c} \leq \min_{\forall n \in \mathcal{N}: d_n^c \neq 0} e_n^c \\
            \overline{p_c} \geq \max_{\forall n \in \mathcal{N}: d_n^c \neq 0} e_n^c.
        \end{array}
    \right.
    \label{p}
\end{equation}
The lower bound means that the MVNO will reject the tenant $n$ directly if $\exists c \in \mathcal{C}, e_n^c$ is lower than $\underline{p_c}$. 
The role of the lower bound is to avoid the tenants deliberately overstating their resource demands to get a discount. In other words, the 
tenants are forced to engage the transactions with their true preferences and no resource will be wasted. The upper bound in \eqref{p} 
is to eliminate irrational tenants or mock auctions, which exists naturally in a wholesome market.

For each tenant $n \in \mathcal{N}$, we use $x_n \in \{0, 1\}$ to indicate whether the deal is successful. The utility of tenant $n$ is 
defined as $U_n \triangleq (v_n - \pi_n) \cdot x_n$, where $\pi_n$ is the payment. The utility of the MVNO is defined as 
$U_o \triangleq \sum_{n \in \mathcal{N}} \pi_n \cdot x_n - \sum_{c \in \mathcal{C}} f_c\Big(\sum_{n \in \mathcal{N}} d_n^c x_n \Big)$, where 
$\forall c \in \mathcal{C}, f_c: [0, 1] \to \mathbb{R}$ is a non-decreasing zero-startup cost function of resource $c$. We set $f_c$ as a non-decreasing 
function because more resources virtualized and sliced, more operating and maintenance costs on those rent-out slices.

In the above formulation, we take $\sigma_n$ and the mapping $\{f_s(\gamma, \tau)\}_{\forall s \in \mathcal{S}_n} \to \mathcal{S}_n$ 
as private information of tenant $n$ which should not be accessible to the MVNO and the other tenants. The former comes down to the payment models and 
pricing strategies adopted by tenant $n$ and reveals the purchasing power of it served users. The latter establishes the relationship between data 
flows and their initiators. In online settings, the deals between the MVNO and the tenants are made one-by-one according to the arrival sequence of 
tenants. Our goal is to (approximately) maximize the social welfare of this ecosystem, i.e. the sum of the MVNO's utility and all the tenants', in 
an \textit{online} and \textit{decentralized} setting. Before introducing the online algorithm proposed in this paper, we first formulate 
the global \textit{offline} social welfare maximization problem as follows.
\begin{subequations}
    \begin{equation*}
        \mathcal{P}_1: \max_{\{x_n\}_{\forall n \in \mathcal{N}}} \sum_{n \in \mathcal{N}} v_n x_n - 
        \sum_{c \in \mathcal{C}} f_c \Big( \sum_{n \in \mathcal{N}} d_n^c x_n \Big)
    \end{equation*}
    \begin{equation}
        s.t. \quad \sum_{n \in \mathcal{N}} d_n^c x_n \leq 1, \forall c \in \mathcal{C},
        \label{p1_con1}
    \end{equation}
    \begin{equation}
        \qquad x_n \in \{0, 1\}, \forall n \in \mathcal{N}.
        \label{p1_con2}
    \end{equation}
\end{subequations}
In $\mathcal{P}_1$, $v_n$ is obtained through \eqref{vn}. Even though the problem is hard to solve, it is formulated based on the complete knowledge 
of the ecosystem. In other words, the formulation of $\mathcal{P}_1$ itself is a detriment on privacy. In an online setting, the MVNO should only 
know the setup information $\{ f_c, \underline{p_c}, \overline{p_c} \}_{\forall c \in \mathcal{C}}$ and the attributes defined in the GSTs 
$\{\varrho_s\}_{\forall s \in \mathcal{S}_n, \forall n \in \mathcal{N}}$ handed in by tenants as a priori. It should not know anything 
about the private tuple 
$$
\vec{\theta} \triangleq \Big( \{\sigma_n\}_{\forall n \in \mathcal{N}}, \big\{ \{f_s(\gamma, \tau)\}_{\forall s \in \mathcal{S}_n} \to \mathcal{S}_n\big\}_{\forall n \in \mathcal{N}} \Big)
$$
and the arrival sequence of tenants. In addition, each tenant should know nothing about the other tenants at all. As a result, to solve the problem 
in a privacy-preserving decentralized setting, we need to ensure that the deal is made with only a small flow of information transferred between 
the MVNO and each tenant without revealing any sensitive information. 
In the proposed algorithm \textsf{DPoS}, which will be introduced in the following, each time when a new tenant $n$ arrives, 
the tenant makes the decision $x_n$ \textit{by itself} according to the disclosed information such as current rental price of each kind resource. 
If $x_n$ is set as $1$, then tenant $n$ sends $(1, \pi_n, \{d_n^c\}_{\forall c \in \mathcal{C}})$ to the MVNO. Otherwise $(0, 0, 0)$ is sent. 
The MVNO can only access the data transferred to it.

\section{Algorithm Design}\label{sec3}
To maximize the social welfare in an online and decentralized setting, we first introduce some notations, then demonstrate the designing of the 
\textbf{D}istributed \textbf{P}rivacy-preserving \textbf{o}nline \textbf{S}licing algorithm, \textsf{DPoS}. 

\subsection{The Primal-Dual Approach}\label{sec3.1}
$\forall c \in \mathcal{C}$, we introduce the \textit{extended cost function} $\tilde{f}_c$ as follows.
\begin{equation}
    \tilde{f}_c(y) \triangleq
    \left\{
        \begin{array}{ll}
            f_c(y) & \textrm{if } y \in [0, 1]\\
            +\infty & \textrm{if } y \in (1, +\infty).
        \end{array}
    \right.
    \label{extended_f}
\end{equation}
$\tilde{f}_c$ extends the domain of $f_c$ to $[0, +\infty)$. Correspondingly, we define the \textit{profit function} $F_{p_c}$ of resource $c$ for 
the MVNO as follows:
\begin{equation}
    F_{p_c} (y_c) \triangleq p_c y_c - \tilde{f}_c(y_c), \forall y_c \in [0, +\infty).
    \label{F_c}
\end{equation}
Regarding $y_c$ as the total resource rented out of type $c$ and $p_c$ as the rental price of resource $c$, $F_{p_c} (y_c)$ is the revenue 
obtained by renting out $y_c$ unit of resource $c$ minus the maintainers cost of it. Based on \eqref{F_c}, we denote the \textit{maximum profit} 
$h_c$ of resource $c$ when the rental price is $p_c$ by
\begin{equation}
    h_c (p_c) \triangleq \max_{y_c \geq 0} F_{p_c} (y_c).
    \label{h_c}
\end{equation}

Following the primal-dual approach \cite{online-mechanism}\cite{primal-dual1}, we introduce the \textit{Relaxed Primal Problem} $\mathcal{P}_2$.
\begin{subequations}
    \begin{equation*}
        \mathcal{P}_2: \max_{\vec{x}, \vec{y}} \sum_{n \in \mathcal{N}} \sum_{s \in \mathcal{S}_n} \varrho_s \cdot \sigma_n \Big(f_s(\gamma, \tau)\Big) x_n - 
        \sum_{c \in \mathcal{C}} \tilde{f}_c ( y_c )
    \end{equation*}
    \begin{equation}
        s.t. \quad \sum_{n \in \mathcal{N}} d_n^c x_n \leq y_c, \forall c \in \mathcal{C},
        \label{p2-cons1}
    \end{equation}
    \begin{equation}
        \qquad \vec{x} \leq \mathbf{1}, \vec{x} \geq 0, \vec{y} \geq 0,
        \label{p2-cons2}
    \end{equation}
\end{subequations}
where $\vec{x} = [x_n]_{n \in \mathcal{N}} \in [0, 1]^N$, and $\vec{y} = [y_c]_{y \in \mathcal{C}} \in \mathbb{R}^C$.
In terms of the relation between $\mathcal{P}_1$ and $\mathcal{P}_2$, we have the following proposition.
\begin{proposition}
    $\mathcal{P}_2$ is equivalent to $\mathcal{P}_1$ except the relaxation of $\{x_n\}_{\forall n \in \mathcal{N}}$.
    \begin{proof}
        To maximize the objective of $\mathcal{P}_1$, the optimal $\vec{y}^{\star}$ must be located in $[0, 1]^N$. Because $f_c$ is 
        non-decreasing for all kinds of resource $c \in \mathcal{C}$, the optimal $y_c^{\star}$ must be the minimum allowed, i.e. 
        $\sum_{n \in \mathcal{N}} d_n^c x_n$. Thus, except relaxing $\{x_n\}_{\forall n \in \mathcal{N}}$ to the continuous interval $[0, 1]^N$, 
        $\mathcal{P}_2$ is the same as $\mathcal{P}_1$.
    \end{proof}
\end{proposition}

Take $\mathcal{P}_2$ as the primal problem, the following proposition gives the dual problem $\mathcal{P}_3$. 
\begin{proposition}
    The dual problem of $\mathcal{P}_2$ is:
    \begin{subequations}
        \begin{equation}
            \mathcal{P}_3: \min_{\vec{p}, \vec{\psi}} \sum_{n \in \mathcal{N}} \psi_n + 
            \sum_{c \in \mathcal{C}} h_c ( p_c )
        \end{equation}
        \begin{equation}
            s.t. \quad \psi_n \geq v_n - \sum_{c \in \mathcal{C}} p_c d_n^c, \forall n \in \mathcal{N},
            \label{dual1}
        \end{equation}
        \begin{equation}
            \qquad \vec{\psi} \geq \mathbf{0}, \vec{p} \geq \mathbf{0},
            \label{dual2}
        \end{equation}
    \end{subequations}
    where $\vec{\psi} = [\psi_n]_{n \in \mathcal{N}} \in \mathbb{R}^N$ and $\vec{p} = [p_c]_{c \in \mathcal{C}} \in \mathbb{R}^C$ are the dual variables 
    corresponding to $\vec{x}$ and $\vec{y}$, respectively.
    \begin{proof}
        By introducing the Lagrangian multipliers $\{p_c\}_{\forall c \in \mathcal{C}}$ and $\{\psi_n\}_{\forall n \in \mathcal{N}}$ for 
        \eqref{p2-cons1} and the first inequality of \eqref{p2-cons2}, respectively, the Lagrangian of $\mathcal{P}_2$ is 
        \begin{eqnarray*}
            \qquad \Lambda (\vec{x}, \vec{y}, \vec{\psi}, \vec{p}) = 
            \sum_{c \in \mathcal{C}} \Big( p_c y_c - \tilde{f}_c (y_c) \Big) + \sum_{n \in \mathcal{N}} \psi_n \quad \\
            \qquad + \sum_{n \in \mathcal{N}} x_n \Bigg( \sum_{s \in \mathcal{S}_n} \varrho_s \cdot \sigma_n \Big(f_s(\gamma, \tau)\Big) - \sum_{c \in \mathcal{C}} p_c d_n^c - \psi_n \Bigg)
        \end{eqnarray*}
        Thus, we have 
        \begin{equation*}
            \min_{\vec{\psi}, \vec{p}} \max_{\vec{x}, \vec{y}} \Lambda = 
        \min_{\vec{\psi}, \vec{p}} \bigg( \max_{\vec{y}} \sum_{c \in \mathcal{C}} \Big( p_c y_c - \tilde{f}_c (y_c) \Big) + \sum_{n \in \mathcal{N}} \psi_n \bigg)
        \end{equation*}
        when $\forall n \in \mathcal{N}$, $\psi_n \geq v_n - \sum_{c \in \mathcal{C}} p_c d_n^c$. Therein, 
        $\sum_{s \in \mathcal{S}_n} \varrho_s \cdot \sigma_n \Big(f_s(\gamma, \tau) \Big)$ is replaced by $v_n$ through \eqref{vn}. The result 
        is immediate with \eqref{h_c}.
    \end{proof}
\end{proposition}
Regarding $\psi_n$ as the utility of tenant $n$. The objective of $\mathcal{P}_3$ is the aggregate utilities of all tenants plus the 
optimal utility of the MVNO. Both the objective of $\mathcal{P}_1$ and $\mathcal{P}_3$ indicate the social welfare of the ecosystem.

\subsection{The \textsf{DPoS} Algorithm}
Note that the rental price $p_c$ of resource $c$ is a global variable known to all tenants. Thus, if the final optimal price $\vec{p}$ is 
known to the MVNO, each time a tenant $n$ arrives, then this tenant can make the rent decision $x_n$ without worrying about whether the optimal social 
welfare is achieved or not. However, it is impossible to know the exact value of $\vec{p}$ in advance without the arrival sequence and 
$\vec{\theta}$. To tackle with this problem, inspired by \cite{primal-dual2}, \cite{online-mechanism} and \cite{primal-dual1}, 
we design the \textsf{DPoS} algorithm based on the alternating update of primal \& dual variables (of $\mathcal{P}_2$ and $\mathcal{P}_3$) and the 
\textit{predict-and-update} of $\vec{p}$. In the following, we place a hat on top of variables that denote the decisions made online.

\begin{figure}[htbp]
    \centerline{\includegraphics[width=3in]{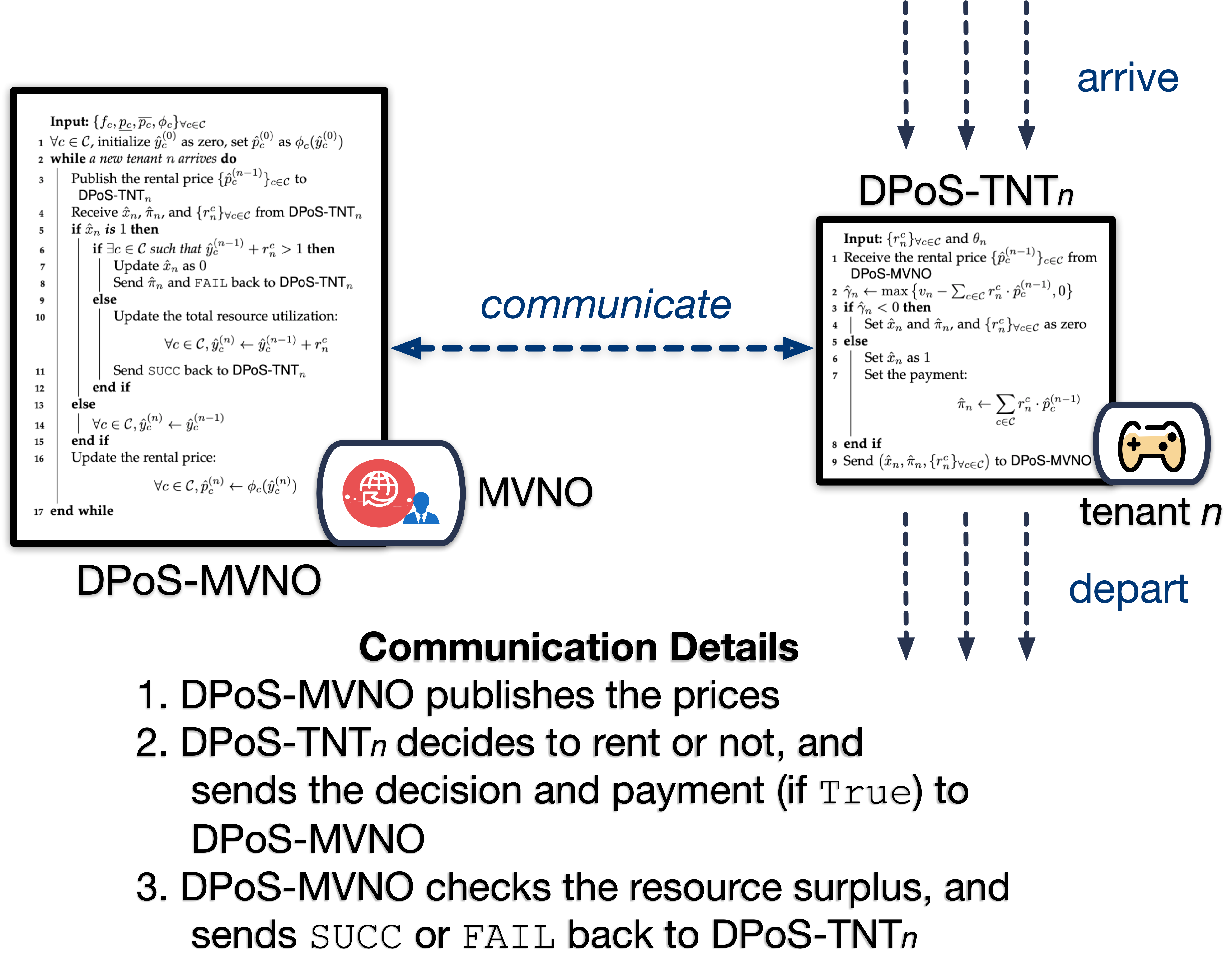}}
    \caption{How \textsf{DPoS} works. Each time a new tenant $n$ arrives, only a small flow of privacy-irrelevant data are transferred between 
    \textsf{DPoS-MVNO} and \textsf{DPoS-TNT}$_n$.}
    \label{fig2}
\end{figure}

\begin{figure}[!ht]
    \removelatexerror
    \begin{algorithm}[H]
        \caption{\textsf{DPoS-MVNO}}
        \KwIn{$\{ f_c, \underline{p_c}, \overline{p_c}, \phi_c \}_{\forall c \in \mathcal{C}}$}
        $\forall c \in \mathcal{C}$, initialize $\hat{y}_c^{(0)}$ as zero, set $\hat{p}_c^{(0)}$ as $\phi_c(\hat{y}_c^{(0)})$ \\
        \While{a new tenant $n$ arrives}
        {
            Publish the rental price $\{ \hat{p}_c^{(n-1)} \}_{c \in \mathcal{C}}$ to \textsf{DPoS-TNT}$_n$\\
            Receive $\hat{x}_n$, $\hat{\pi}_n$, and $\{d_n^c\}_{\forall c \in \mathcal{C}}$ from \textsf{DPoS-TNT}$_n$\\
            \uIf{$\hat{x}_n$ \textbf{is} $1$}
            {
                \uIf{$\exists c \in \mathcal{C}$ such that $\hat{y}_c^{(n-1)} + d_n^c > 1$}
                {
                    Update $\hat{x}_n$ as $0$\\
                    Send $\hat{\pi}_n$ and \texttt{FAIL} back to \textsf{DPoS-TNT}$_n$
                }
                \Else
                {
                    Update the total resource utilization: $$\forall c \in \mathcal{C}, \hat{y}_c^{(n)} \leftarrow \hat{y}_c^{(n-1)} + d_n^c$$\\
                    Send \texttt{SUCC} back to \textsf{DPoS-TNT}$_n$
                }
            }
            \Else
            {
                $\forall c \in \mathcal{C}, \hat{y}_c^{(n)} \leftarrow \hat{y}_c^{(n-1)}$\\
            }
            Update the rental price: $$ \forall c \in \mathcal{C}, \hat{p}_c^{(n)} \leftarrow \phi_c (\hat{y}_c^{(n)}) $$
        }
    \end{algorithm}
\end{figure}

\textsf{DPoS} consists of two parts, \textsf{DPoS-MVNO} and \textsf{DPoS-TNT}$_n$ (each for a tenant). Before a new tenant $n$ arrives, 
\textsf{DPoS-MVNO} prices for each resource $c$ with a function $\phi_c$:
\begin{equation}
    \hat{p}_c^{(n-1)} = \phi_c (\hat{y}_c^{(n-1)}), \forall c \in \mathcal{C}.
    \label{price}
\end{equation}
The pricing functions $\{\phi_c\}_{c \in \mathcal{C}}$ are closely associated to the properties of the cost functions $\{f_c\}_{c \in \mathcal{C}}$. 
We will provide the analytic forms of them in the follwing subsection.

\textsf{DPoS-MVNO} discloses the rental prices $\{ \hat{p}_c^{(n-1)} \}_{c \in \mathcal{C}}$ to tenant $n$. Then, tenant $n$ judges whether it has 
\textit{positive} utility if it decides to rent $\{d_n^c\}_{\forall c \in \mathcal{C}}$ ($\hat{x}_n \leftarrow 1$). If yes, \textsf{DPoS-TNT}$_n$ 
sets the payment $\hat{\pi}_n$ as $\sum_{c \in \mathcal{C}} d_n^c \cdot \hat{p}_c^{(n-1)}$. Otherwise, both $\hat{x}_n$ and $\hat{\pi}_n$ are set 
as zero. In the end, \textsf{DPoS-TNT}$_n$ sends $(\hat{x}_n, \hat{\pi}_n, \{d_n^c\}_{\forall c \in \mathcal{C}})$ to \textsf{DPoS-MVNO}.

When \textsf{DPoS-MVNO} receives the message from \textsf{DPoS-TNT}$_n$, it checks whether the resource surplus can satisfy tenant $n$. If yes, 
\textsf{DPoS-MVNO} sends the indicator \texttt{SUCC} to \textsf{DPoS-TNT}$_n$ to inform the success of this transaction. Otherwise, it sends \texttt{FAIL} 
and returns the rent $\hat{\pi}_n$. If succeed, tenant $n$ hands in the GST and other matters that need to be provided.

The procedure is visualized in Fig. \ref{fig2}.
Note that the data transfer between \textsf{DPoS-MVNO} and \textsf{DPoS-TNT}$_n$ is \textit{stop-and-wait}, i.e., a new arrival tenant will not be 
handed by the MVNO until the transaction between the MVNO and the previous tenant is done.
In \textsf{DPoS}, only a small flow of privacy-irrelevent data $(\hat{x}_n, \hat{\pi}_n, \{d_n^c\}_{\forall c \in \mathcal{C}})$ are 
transferred between \textsf{DPoS-TNT}$_n$ and \textsf{DPoS-MVNO}. The MVNO cannot collect any information from $\vec{\theta}$. In addition, each tenant 
knows nothing about the other tenants. \textsf{DPoS} is implemented in a posted price manner \cite{posted1}\cite{posted2}, where the rent decision 
made by each tenant is only \textit{take-it-or-leave-it}. A tenant cannot get any discount even if it rents relatively large amounts of resources, 
which leads to the fact that \textit{how much to use, how much to rent.} No resource will be wasted.

It is easy to verify that the complexity of \textsf{DPoS} is linear with the scale of tenants and type of resources.
In \textsf{DPoS-MVNO}, the while-loop terminates after all the $|\mathcal{N}|$ tenants finish their transactions in turn. During 
the loop, the most time-consuming operation lies in step 6 and step 10, where the MVNO needs check whether each type of resource $c$ is enough to 
support a transaction and take $d_n^c$ off if permitted. In worst case, the number of operations is $2 |\mathcal{C}|$. Considering that all the 
left steps can be executed in $O(1)$-complexity, the worst-case complexity of \textsf{DPoS-MVNO} is $O(|\mathcal{N}| \cdot |\mathcal{C}|)$. 
As for \textsf{DPoS-TNT}$_n$, time-consumption operations lie in step 2, step 4, and step 7, all of which are $O(|\mathcal{C}|)$-complexity. 
Therefore, \textsf{DPoS-TNT}$_n$ is of $O(|\mathcal{C}|)$-complexity in worst case.

\subsection{The Dynamic Pricing Functions}
In \textsf{DPoS}, the only difficulty lies in that how the pricing functions $\{\phi_c\}_{c \in \mathcal{C}}$ are designed. As mentioned 
before, the analytic forms of $\{\phi_c\}_{c \in \mathcal{C}}$ strongly rely on the properties of cost functions $\{f_c\}_{c \in \mathcal{C}}$. 
Even so, we claim that \textit{in \textsf{DPoS}, $\{\phi_c\}_{\forall c \in \mathcal{C}}$ are monotonically non-decreasing positive functions}.
We set $\phi_c$ as a non-decreasing function because it profoundly reflects the underlying economic phenomenon, i.e., \textit{a thing is valued 
in proportion to its rarity.} The later the tenant comes to renting the remaining resources, the higher cost it has to pay.

\begin{figure}[!ht]
    \removelatexerror
    \begin{algorithm}[H]
        \caption{\textsf{DPoS-TNT}$_n$}
        \KwIn{$\{d_n^c\}_{\forall c \in \mathcal{C}}$ and $\theta_n$}
        Receive the rental price $\{ \hat{p}_c^{(n-1)} \}_{c \in \mathcal{C}}$ from \textsf{DPoS-MVNO}\\
        $\hat{\psi}_n \leftarrow \max \big\{ v_n - \sum_{c \in \mathcal{C}} d_n^c \cdot \hat{p}_c^{(n-1)}, 0 \big\}$\\
        \uIf{$\hat{\psi}_n < 0$}
        {
            Set $\hat{x}_n$ and $\hat{\pi}_n$, and $\{d_n^c\}_{\forall c \in \mathcal{C}}$ as zero\\
        }
        \Else
        {
            Set $\hat{x}_n$ as $1$\\
            Set the payment:
            $$ \hat{\pi}_n  \leftarrow \sum_{c \in \mathcal{C}} d_n^c \cdot \hat{p}_c^{(n-1)} $$
        }
        Send $\big(\hat{x}_n, \hat{\pi}_n, \{d_n^c\}_{\forall c \in \mathcal{C}}\big)$ to \textsf{DPoS-MVNO}
    \end{algorithm}
\end{figure}

Now, we demonstrate the forms of $\{\phi_c\}_{\forall c \in \mathcal{C}}$ when the costs are linear. 
Concretely, if $\forall c \in \mathcal{C}$, the cost function has the form 
\begin{equation}
    f_c (y) = q_c y,
    \label{linear_cost}
\end{equation}
where $0 < q_c < \underline{p_c}$. Then, in \textsf{DPoS}, the pricing function $\phi_c$ is set as follows:
\begin{eqnarray}
    \phi_c (y) = 
    \left\{
        \begin{array}{ll}
            \underline{p_c} & y \in [0, w_c) \\
            q_c + (\underline{p_c} - q_c) \cdot e^{y/w_c - 1} & y \in [w_c, 1] \\
            +\infty & y \in (1, +\infty),
        \end{array}
    \right.
    \label{linear_price}
\end{eqnarray}
where
\begin{equation}
    w_c = \bigg( 1 + \ln \frac{\sum_{c' \in \mathcal{C}} (\overline{p_{c'}} - q_{c'})}{\underline{p_c} - q_c} \bigg)^{-1}
    \label{w_c}
\end{equation}
is a threshold. Tan et al. also discuss the construction of the pricing function (for single resource and multiple substitutable resources) when the resource's cost 
function is strictly-convex \cite{online-mechanism}, which involves the solving of several first-order two-point boundary value problems (BVPs) 
\cite{ODE}. In the next section, we will show that the competitive ratio of \textsf{DPoS} is the optimal one over all the online algorithms 
when $\{f_c\}_{c \in \mathcal{C}}$ are linear.

\section{Theoretical Analysis}\label{sec4}
The commonly used measure for online algorithms is the standard competitive 
analysis framework \cite{competitive}. The definition of \textit{competitive ratio} for any online algorithm to $\mathcal{P}_1$ is given below.
\begin{definition}
    For any arrival instance $1, 2, ..., N$, denoted by $\mathcal{A}$, the competitive ratio for an online algorithm is defined as 
    \begin{equation}
        \alpha \triangleq \max_{\forall \mathcal{A}} \frac{\mathbf{\Theta}_{\text{off}} (\mathcal{A})}{\mathbf{\Theta}_{\text{on}} (\mathcal{A})},
    \end{equation}
    where $\mathbf{\Theta}_{\text{off}} (\mathcal{A})$ is the maximum objective value of $\mathcal{P}_1$, $\mathbf{\Theta}_{\text{on}} (\mathcal{A})$ 
    is the objective function value of $\mathcal{P}_1$ obtained by this online algorithm. 
\end{definition}
Obviously, $\alpha \geq 1$ always holds. The smaller $\alpha$ is, the better the online algorithm. An online algorithm is \textit{competitive} 
if its competitive ratio is upper bounded. Further, we can define the optimal competitive ratio as 
\begin{equation}
    \alpha^\star \triangleq \inf \max_{\forall \mathcal{A}} \frac{\mathbf{\Theta}_{\text{\textit{off}}} (\mathcal{A})}{\mathbf{\Theta}_{\text{\textit{on}}} (\mathcal{A})},
\end{equation}
where the $\inf$ is taken w.r.t. all possible online algorithms. In the following, we drop the parenthesis and $\mathcal{A}$ for simplification. 
Note that whether optimal or not, competitive ratio only gives the \textit{worst-case} guarantee.

To analyze the competitive ratio achieved by \textsf{DPoS}, we need to introduce several propositions and theorems beforehand. We will firstly 
verify that \textsf{DPoS} is $\alpha$-competitive for some constant $\alpha$, then prove that it is the optimal one over all online 
algorithms when $\{f_c\}_{\forall c \in \mathcal{C}}$ are linear. The first proposition introduced is related to the maximum utility $h_c$.
\begin{proposition}
    $\forall c \in \mathcal{C}$, the function $h_c$, defined in \eqref{h_c}, can also be written as
    \begin{equation}
        h_c (p_c) = 
        \left\{
            \begin{array}{ll}
                F_{p_c} \big(f_c'^{-1}(p_c)\big) & p_c \in [\underline{\varsigma_c}, \overline{\varsigma_c}] \\
                F_{p_c} (1) & p_c \in (\overline{\varsigma_c}, +\infty), \\
            \end{array}
        \right.
        \label{conjugate}
    \end{equation}
    where $\underline{\varsigma_c} \triangleq f_c'(0)$, $\overline{\varsigma_c} \triangleq f_c'(1)$, $f_c'$ is the derivative of 
    $f_c$, and $f_c'^{-1}$ is the inverse of $f_c'$.
    \begin{proof}
        $\forall c \in \mathcal{C}$, when $\underline{\varsigma_c} \leq p_c \leq \overline{\varsigma_c}$, regarding $p_c$ as the derivative of 
        the non-decreasing $f_c$, then we have $f_c'^{-1}(p_c) \in [0, 1]$. Now we need to find the $y_c^\star$ which maximizes $F_{p_c} (y_c)$. By analyzing 
        the property of $\partial F_{p_c} (y_c) / \partial y_c$, which is $p_c - f_c'(y_c)$, we can find that the exact $y_c^\star$ satisfies 
        $p_c = f'(y_c^\star)$. Thus $h_c (p_c)$ is $F_{p_c} \big(f_c'^{-1}(p_c)\big)$ when $\underline{\varsigma_c} \leq p_c \leq \overline{\varsigma_c}$. 
        The same applies to the second segment of \eqref{conjugate}. 
    \end{proof}
\end{proposition}
\eqref{conjugate} is known as the \textit{convex conjugate} of $\tilde{f}_c$ \cite{convex-opt}. For a given online algorithm, denote the objective 
of $\mathcal{P}_2$ and $\mathcal{P}_3$ by $\mathbf{\Theta}_{\mathcal{P}_2}^n $ and $\mathbf{\Theta}_{\mathcal{P}_3}^n$ after processing tenant $n$, 
respectively. Also, we use $\mathcal{V}_{\mathcal{P}_2} \triangleq \big(\{ \hat{x}_n \}_{\forall n \in \mathcal{N}}, \hat{y}_N\big)$ and 
$\mathcal{V}_{\mathcal{P}_3} \triangleq \big(\{ \hat{\psi}_n \}_{\forall n \in \mathcal{N}}, \hat{p}_N\big)$ to denote the complete set of online 
primal and dual solutions, respectively. In the following, we demonstrate the sufficient conditions of designing an $\alpha$-competitive online 
algorithm for $\mathcal{P}_1$, and then show that \textsf{DPoS} satisfies the conditions. 
\begin{proposition}
    (Adapted from proposition 3.1 of \cite{online-mechanism}) When $\{f_c\}_\forall c \in \mathcal{C}$ are linear\footnote{Proposition 1 of 
    \cite{online-mechanism} also requires that $\{\underline{\varsigma_c} < \underline{p_c}\}_{\forall c \in \mathcal{C}}$ holds, which is not 
    required in this proposition.}, an online algorithm is $\alpha$-competitive if the following conditions are satisfied:
    \begin{itemize}
        \item All the online primal solutions in $\mathcal{V}_{\mathcal{P}_2}$ are feasible to $\mathcal{P}_1$;
        \item All the online dual solutions in $\mathcal{V}_{\mathcal{P}_3}$ are feasible to $\mathcal{P}_3$;
        \item There exists a tenant $k \in \mathcal{N}$ such that 
        \begin{equation}
            \mathbf{\Theta}_{\mathcal{P}_2}^k \geq \frac{1}{\alpha} \mathbf{\Theta}_{\mathcal{P}_3}^k
            \label{con0}
        \end{equation}
        and $\forall n \in \{k+1, ..., N\}$,
        \begin{equation}
            \mathbf{\Theta}_{\mathcal{P}_2}^n  - \mathbf{\Theta}_{\mathcal{P}_2}^{n-1}  \geq \frac{1}{\alpha} 
            \big( \mathbf{\Theta}_{\mathcal{P}_3}^n - \mathbf{\Theta}_{\mathcal{P}_3}^{n-1} \big)
            \label{con1}
        \end{equation}
        holds.
    \end{itemize}
\end{proposition}
\begin{proof}
    Let us denote the optimal objective of $\mathcal{P}_2$ and $\mathcal{P}_3$ as $\mathbf{\Theta}_{\mathcal{P}_2}^\star$ and 
    $\mathbf{\Theta}_{\mathcal{P}_3}^\star$, respectively. Then, 
    \begin{equation}
        \mathbf{\Theta}_{\text{\textit{off}}} \leq \mathbf{\Theta}_{\mathcal{P}_2}^\star = \mathbf{\Theta}_{\mathcal{P}_3}^\star 
        \leq \mathbf{\Theta}_{\mathcal{P}_3}^N.
    \end{equation}
    The reason for the first inequality is that $\mathcal{P}_2$ is a relaxation of $\mathcal{P}_1$. The reason for the first equality is that 
    when $\{f_c\}_{\forall c \in \mathcal{C}}$ are linear, strong duality holds between $\mathcal{P}_2$ and $\mathcal{P}_3$.
    Besides, $\mathbf{\Theta}_{\text{\textit{on}}} = \mathbf{\Theta}_{\mathcal{P}_2}^N$. As a result, to make 
    $\alpha \geq \mathbf{\Theta}_{\text{\textit{off}}}/\mathbf{\Theta}_{\text{\textit{on}}}$ always hods, we can try to ensure that 
    $\mathbf{\Theta}_{\mathcal{P}_2}^N \geq \frac{1}{\alpha} \mathbf{\Theta}_{\mathcal{P}_3}^N$ holds. 

    According to \eqref{con1}, the following inequalities holds:
    \begin{eqnarray*}
        \sum_{n \in \mathcal{N}, n > k} \Big(\mathbf{\Theta}_{\mathcal{P}_2}^n - \mathbf{\Theta}_{\mathcal{P}_2}^{n-1}\Big) &\geq& 
        \frac{1}{\alpha} \sum_{n \in \mathcal{N}, n > k} \Big(\mathbf{\Theta}_{\mathcal{P}_3}^n  - \mathbf{\Theta}_{\mathcal{P}_3}^{n-1}\Big) \\
        \Longleftrightarrow \qquad \mathbf{\Theta}_{\mathcal{P}_2}^N  - \mathbf{\Theta}_{\mathcal{P}_2}^k &\geq& 
        \frac{1}{\alpha} \Big( \mathbf{\Theta}_{\mathcal{P}_3}^N  - \mathbf{\Theta}_{\mathcal{P}_3}^k \Big) \\
        \Longleftrightarrow \quad \mathbf{\Theta}_{\mathcal{P}_2}^N - \frac{1}{\alpha} \mathbf{\Theta}_{\mathcal{P}_3}^N &\geq& 
        \mathbf{\Theta}_{\mathcal{P}_2}^k - \frac{1}{\alpha} \mathbf{\Theta}_{\mathcal{P}_3}^k \\
        \Longleftrightarrow \qquad \mathbf{\Theta}_{\mathcal{P}_2}^N &\geq& \frac{1}{\alpha} \mathbf{\Theta}_{\mathcal{P}_3}^N. \qquad \vartriangleright \eqref{con0}
    \end{eqnarray*}
    We thus complete the proof.
\end{proof}
Proposition 4 gives three conditions for designing an $\alpha$-competitive online algorithm when $\{f_c\}_\forall c \in \mathcal{C}$ are linear. 
If we can prove that these conditions hold for \textsf{DPoS}, then we prove that \textsf{DPoS} is at least $\alpha$-competitive for some constant $\alpha$. 
In the following, we prove that the first and the second condition hold. 
\begin{itemize}
    \item It is obvious that $\mathcal{V}_{\mathcal{P}_2}$ obtained by \textsf{DPoS} is feasible to $\mathcal{P}_2$ because the 
    ``\textbf{if statement}'' in step 6 of \textsf{DPoS-MVNO} and step 4 \& 6 of \textsf{DPoS-TNT}$_n$ ensure that \eqref{p1_con1} and 
    \eqref{p1_con2} can never be violated.
    \item From step 2 of \textsf{DPoS-TNT}$_n$ we can find that 
    $\forall c \in \mathcal{C}, \hat{\psi}_n \geq v_n - \sum_{c \in \mathcal{C}} d_n^c \cdot \hat{p}_c^{(n-1)}$. 
    Because $\{\phi_c\}_{\forall c \in \mathcal{C}}$ defined in \textsf{DPoS} are non-decreasing positive functions, the following inequality
    $$\hat{p}_c^{(N)} \geq \hat{p}_c^{(n)} \geq \hat{p}_c^{(n-1)} > 0$$
    holds. Thus $\forall n \in \mathcal{N}$, $\hat{\psi}_n \geq v_n - \sum_{c \in \mathcal{C}} d_n^c \hat{p}_c^{(N)}$ holds, 
    where $\hat{p}_c^{(N)}$ is the final rental price of resource $c$, i.e. $p_c$ in $\mathcal{P}_3$. 
    Thus, \eqref{dual1} not violated. Step 2 of \textsf{DPoS-TNT}$_n$ ensures that $\hat{\vec{\psi}} \geq \vec{0}$ holds. 
    Also note that in \textsf{DPoS} $\{\phi_c\}_{\forall c \in \mathcal{C}}$ are non-decreasing positive functions, which leads to 
    $\hat{\vec{p}} \geq \vec{0}$ always holds. We thus prove that \eqref{dual2} is not violated. Since both \eqref{dual1} and \eqref{dual2} 
    are not violated, the second condition in proposition 4 holds for \textsf{DPoS}.
\end{itemize}
The proof of that the third condition holds is related to the design of the pricing functions $\{\phi_c\}_{\forall c \in \mathcal{C}}$. 
The following theorem shows that when $\{\phi_c\}_{\forall c \in \mathcal{C}}$ in \textsf{DPoS} are designed as \eqref{41} $\sim$ \eqref{43} 
indicate, the third condition in proposition 4 holds.

\begin{theorem}
    (Adapted from theorem 4.1 of \cite{online-mechanism}) When $\{f_c\}_{\forall c \in \mathcal{C}}$ are linear and 
    $\{0 < \underline{\varsigma_c} < \underline{p_c}\}_{\forall c \in \mathcal{C}}$ holds, if $\forall c \in \mathcal{C}$, the pricing function 
    $\phi_c$ in \textsf{DPoS} has the form: 
    \begin{equation}
        \phi_c (y) = 
        \left\{
            \begin{array}{ll}
                \underline{p_c} & y \in [0, w_c) \\
                \varphi_c (y) & y \in [w_c, 1] \\
                +\infty & y \in (1, +\infty),
            \end{array}
        \right.
        \label{41}
    \end{equation}
    where 
    \begin{equation}
        w_c \in \Big[0, \argmax_{y \geq 0} \underline{p_c} y - \tilde{f}_c(y)\Big]
        \label{41-add}
    \end{equation}
    is a threshold that satisfies
    \begin{equation}
        F_{\underline{p_c}} (w_c) \geq \frac{1}{\alpha_c} h_c (\underline{p_c}),
        \label{42}
    \end{equation}
    and $\varphi_c (y)$ is an increasing function that satisfies 
    \begin{equation}
        \left\{
        \begin{array}{l}
            \varphi'_c (y) \leq \alpha_c \cdot \frac{\varphi_c (y) - f'_c(y)}{h_c'(\varphi_c (y))}, \text{if } y \in (w_c, 1) \\
            \varphi_c (w_c) = \underline{p_c} \\
            \varphi_c (1) \geq \overline{p_c} + \sum_{c' \in \mathcal{C}\backslash\{c\}} h_{c'} (\overline{p_{c'}}),
        \end{array}
        \right.
        \label{43}
    \end{equation}
    then \textsf{DPoS} is $\max_{c \in \mathcal{C}} \alpha_c$-competitive.
\end{theorem}
\begin{proof}
Assume that $\forall c \in \mathcal{C}, w_c = \sum_{n=1}^k d_n^c$, which means that $k$ is the number of tenants such that the total resource 
rented out of type $c$ equals $w_c$. Substitute the definition of $F_{p_c}(\cdot)$ into \eqref{42}, we have
\begin{equation*}
    \underline{p_c} \cdot \Big( \sum_{n=1}^k d_n^c \Big) - \tilde{f}_c \Big( \sum_{n=1}^k d_n^c \Big) \geq \frac{1}{\alpha_c} h_c (\underline{p_c}).
\end{equation*}
Because $\alpha_c \geq 1$ holds for each $c \in \mathcal{C}$ and $\hat{\vec{\psi}} \geq \vec{0}$, the above inequality leads to
\begin{eqnarray*}
    \Big( 1 - \frac{1}{\alpha_c} \Big) \sum_{n=1}^k \hat{\psi}_n 
    &+& \sum_{c \in \mathcal{C}} \Bigg( 
        \underline{p_c} \cdot \Big( \sum_{n=1}^k d_n^c \Big) - \tilde{f}_c \Big( \sum_{n=1}^k d_n^c \Big) 
    \Bigg) \\
    &\geq& \sum_{c \in \mathcal{C}} \frac{1}{\alpha_c} h_c (\underline{p_c}).
\end{eqnarray*}
Further, we have 
\begin{eqnarray}
    &\sum_{n=1}^k \Big(\hat{\psi}_n + \sum_{c \in \mathcal{C}} \underline{p_c} \cdot d_n^c\Big) - \sum_{c \in \mathcal{C}} \tilde{f}_c \Big( \sum_{n=1}^k d_n^c \Big) \nonumber\\
    &\geq \min_{c' \in \mathcal{C}} \frac{1}{\alpha_{c'}} \bigg(\sum_{n=1}^k \hat{\psi}_n  + \sum_{c \in \mathcal{C}} h_c (\underline{p_c}) \bigg).
    \label{prove-t1-1}
\end{eqnarray}
The pricing function in \eqref{41} indicates that the requirements of all tenants will be satisfied as long as each resource $c$'s utilization
is below $w_c$. Thus, we have $\hat{y}_c^{(k)} = \sum_{n=1}^k d_n^c = w_c$. 
Besides, the rental price of resource $c$ these tenants experienced is the same, i.e., $\underline{p_c}$. 
Therefore, \eqref{prove-t1-1} indicates 
$\mathbf{\Theta}_{\mathcal{P}_2}^k \geq \min_{c \in \mathcal{C}} \frac{1}{\alpha_{c}} \mathbf{\Theta}_{\mathcal{P}_3}^k$. 
Meanwhile, it is obvious that $w_c$ must be less than or equal to $\argmax_{y \geq 0} \underline{p_c} y - \tilde{f}_c(y)$ because the rental 
price must be larger than or equal to the \textit{marginal cost} $f_c'(w_c)$ (the result is immediate with \eqref{conjugate}). 

The above has proved that \eqref{con0} holds. In the following, we prove \eqref{con1} holds. The change in the objective of 
$\mathcal{P}_2$ when a new tenant $n$ arrives is
\begin{eqnarray*}
    \mathbf{\Theta}_{\mathcal{P}_2}^n - \mathbf{\Theta}_{\mathcal{P}_2}^{n-1} &=& \hat{\psi}_n + \sum_{c \in \mathcal{C}} \phi_c (\hat{y}_c^{(n-1)}) 
    \Big( \hat{y}_c^{(n)} - \hat{y}_c^{(n-1)} \Big) \\
    &-& \sum_{c \in \mathcal{C}} \Big( \tilde{f}_c (\hat{y}_c^{(n)}) - \tilde{f}_c (\hat{y}_c^{(n-1)}) \Big).
\end{eqnarray*}
The change in the objective of $\mathcal{P}_3$ when a new tenant $n$ arrives is
\begin{equation*}
    \mathbf{\Theta}_{\mathcal{P}_3}^n - \mathbf{\Theta}_{\mathcal{P}_3}^{n-1} = 
    \hat{\psi}_n + \sum_{c \in \mathcal{C}} \Big( h_c (\hat{p}_c^{(n)}) - h_c (\hat{p}_c^{(n-1)}) \Big).
\end{equation*}
To guarantee \eqref{con1} holds, it is \textit{equivalent} to guarantee the following \textit{per-resource} inequality
\begin{eqnarray*}
    &\phi_c (\hat{y}_c^{(n-1)}) \Big( \hat{y}_c^{(n)} - \hat{y}_c^{(n-1)} \Big) - 
    \Big( \tilde{f}_c (\hat{y}_c^{(n)}) - \tilde{f}_c (\hat{y}_c^{(n-1)}) \Big) \\ 
    &\geq \frac{1}{\alpha_c} \Big( h_c (\hat{p}_c^{(n)}) - h_c (\hat{p}_c^{(n-1)}) \Big).
\end{eqnarray*}
Divide both side of the above inequality by $\hat{y}_c^{(n)} - \hat{y}_c^{(n-1)}$, we get 
\begin{equation}
    \phi_c (y_c) - \tilde{f}_c'(y_c) \geq \frac{1}{\alpha_c} \cdot h_c'(\phi_c(y_c)) \cdot \phi_c'(y_c)
    \label{prove-t1-2}
\end{equation}
when $y_c \in [w_c, 1)$. This means that if $\forall y_c \in [w_c, 1)$, \eqref{prove-t1-2} holds, the incremental inequality in \eqref{con1} 
holds for all $y_c \in [w_c, 1)$ for each type of resource. This result is exactly the first segment of \eqref{43}. The second segment of \eqref{43} is to ensure 
the continuity of $\phi_c$. The third segment of \eqref{43} is to make up the missing proof for \eqref{con1} on the exact point $y_c = 1$, which 
can be derived by the deformation of 
\begin{equation*}
    \underline{p_c} w_c + \int_{w_c}^1 \phi_c (y_c) d y_c - \tilde{f}_c (1) \geq \frac{1}{\alpha_c} \sum_{c \in \mathcal{C}} h_c (\overline{p_c}).
\end{equation*}
The above inequality is obtained by taking integration of both sides of \eqref{prove-t1-2}.

So far, we have proved that when $\{\phi_c\}_{\forall c \in \mathcal{C}}$ in \textsf{DPoS} are designed as \eqref{41} $\sim$ \eqref{43} suggested, 
the thrid condition in proposition 4, i.e., \eqref{con0} and \eqref{con1} hold. Thus, we have proved that \textsf{DPoS} is 
$\max_{c \in \mathcal{C}} \alpha_{c}$-competitive.
\end{proof}

In the following, we verify that the design of $\{\phi_c\}_{c \in \mathcal{C}}$ in \textsf{DPoS} when $\{f_c\}$ are linear, which is demonstrated 
in \eqref{linear_price}, satisfies the requirements defined in \eqref{41} $\sim$ \eqref{43}.
When $f_c(y) = q_c y$ and $q_c > 0$, the conjugate $h_c (p_c)$ defined in \eqref{h_c} is given by
\begin{equation}
    h_c (p_c) = 
    \left\{
        \begin{array}{ll}
            0 & p_c \in [0, q_c] \\
            p_c - q_c & p_c \in (q_c, +\infty)
        \end{array}
    \right.
\end{equation}
Note that $0 < q_c < \underline{p_c} \leq \overline{p_c}$. In this case, \eqref{42} is equal to 
\begin{equation*}
    \underline{p_c} w_c - f(w_c) \geq \frac{1}{\alpha_c} \Big( \underline{p_c} - f_c (1) \Big),
\end{equation*}
which indicates $w_c \geq \frac{1}{\alpha_c}$. \eqref{43} is thus equal to 
\begin{equation*}
    \left\{
        \begin{array}{l}
            \varphi_c (y) - f'_c(y) \geq \frac{1}{\alpha_c} \cdot  \varphi'_c (y) \cdot h_c'(\varphi_c (y)), w_c < y < 1\\
            \varphi_c (w_c) = \underline{p_c} \\
            \varphi_c (1) = \sum_{c \in \mathcal{C}} \overline{p_c} - \sum_{c' \in \mathcal{C} \backslash \{c\}} q_{c'}.
        \end{array}
        \right.
\end{equation*}
To minimize $\alpha_c$, it suffices to set $w_c$ as $1/\alpha_c$ and thus the above BVP leads to \eqref{linear_price} and \eqref{w_c}. 

The above analysis leads to the following theorem immediately.
\begin{theorem}
    When the cost functions $\{f_c\}_{\forall c \in \mathcal{C}}$ are linear and $\{0 < \underline{\varsigma_c} < \underline{p_c}\}_{\forall c \in \mathcal{C}}$ holds, 
    the competitive ratio $\alpha$ \textsf{DPoS} achieves is the optimal one over all possible online algorithms. Further, its value is 
    \begin{equation}
        \alpha = \max_{\forall c \in \mathcal{C}} \alpha_c
        = \max_{\forall c \in \mathcal{C}} \frac{1}{w_c},
        \label{worst-ratio}
    \end{equation}
    where $w_c$ is defined in \eqref{w_c}.
\end{theorem}

\section{Experimental Results}\label{sec5}
In this section, we conduct extensive simulation experiments to evaluate the effectiveness and efficiency of \textsf{DPoS}. 
We firstly verify the performance of \textsf{DPoS} against several popular algorithms and handcrafted benchmarking policies on social 
welfare, efficiency, and competitive ratio. Then, we analyze the impact of several system parameters. 

We summarize the key findings of our experiments as follows, and details can be found in Sec. \ref{sec5.2} and Sec. \ref{sec5.3}.
\begin{itemize}
    \item \textsf{DPoS} not only achieves the highest social welfare among all the online algorithms compared, but also shows the 
    \textit{close-to-offline-optimal} performance, especially when the number of tenants is not more than $100$ and the number of 
    resource type is $1$.
    \item In most cases, the ratio of the optimal social welfare to the social welfare achieved by \textsf{DPoS} (fluctuate between 
    $1.00$ and $2.57$) is far less than the worst-case guarantee, i.e. the competitive ratio calculated by \eqref{w_c} and 
    \eqref{worst-ratio} (fluctuate between $5.82$ and $8.54$). 
    \item \textsf{DPoS} is insensitive to environment parameters such as the distribution of $\{d_n^c\}_{\forall c \in \mathcal{C}}$ 
    and the value of the coefficient of the linear cost, $\{q_c\}_{\forall c \in \mathcal{C}}$.
    \item \textsf{DPoS} achieves a satisfactory balance between the overheads (corss-agent communication data size, algorithm's 
    running time, etc.) and the performance.
\end{itemize}

\subsection{Experiment Setup}\label{sec5.1}
By default, we set the number of tenants $N$ as $100$. We also set the number of types of resources as $3$ in default because 
the resources can be roughly divided into computation, storage, and forwarding/bandwidth. Note that $100$ and $3$ are only default 
settings. In Sec. \ref{sec5.2} and Sec. \ref{sec5.3}, we will 
analyze the scalability of \textsf{DPoS} extensively. 

For each tenant $n$, $\{d_n^c\}_{\forall c \in \mathcal{C}}$ is uniformly sampled from the Gaussian distribution 
$N (\mu = \frac{1}{N}, \sigma = \frac{1}{N^2})$. 
The pay level $l_n$ is randomly sampled from $[2, 6]$. The highest level of QoS, denoted by $l_n^h$, is randomly sampled from 
$U(2, 6)$. The lowest level of QoS, denoted by $l_n^0$, is \textit{free user level}. We set the percentage of free users near $40\%$ for 
each tenant \cite{data}. Moreover, the remaining users are randomly assigned to a QoS level according to the pyramid structure. The higher 
the QoS level, the fewer the users. The payment of user $s \in \mathcal{S}_n$ is proportional to his QoS level. By default, 
$\forall n \in \mathcal{N}, \forall s \in \mathcal{S}_n$, we set $\sigma_n$ as identity function. For each type of resource, we take linear 
cost defined in \eqref{linear_cost}. By default, $\forall c \in \mathcal{C}$, $q_c$ is randomly chosen from 
$[\frac{1}{6}\underline{p_c}, \frac{5}{6}\underline{p_c}]$.

\begin{table}[htbp]   
    \begin{center}
    \caption{\label{para}Default parameter settings.}   
    \begin{tabular}{c|c|c|c}    
        \toprule
        {\textsf{\textbf{Parameter}}} & {\textsf{\textbf{Value}}} & {\textsf{\textbf{Parameter}}} & {\textsf{\textbf{Value}}}\\[+0.1mm]
        \midrule
        $N$ & $100$ & $C$ & $3$\\[+0.7mm]
        $\{d_n^c\}_{\forall c \in \mathcal{C}}$ & $\sim N (\mu = \frac{1}{N}, \sigma = \frac{1}{N^2})$ & $l_n^h$ & $\sim U(2, 6)$ \\[+0.7mm]
        $\mathcal{S}_n$ & $\sim N (\mu = 10^6, \sigma = 10^5)$ & $\Pr (l_n^0)$ & $\approx 40\%$ \\[+0.7mm]
        $q_c$ & $\sim U(\frac{1}{6}\underline{p_c}, \frac{5}{6}\underline{p_c})$ &  $\sigma_n$ & identity\\[+0.7mm]
        \bottomrule   
    \end{tabular}  
    \end{center}
\end{table}
\textsf{DPoS} is compared with the following algorithms. Thereinto, \textit{CVX} and \textit{Heuristic} are used to obtain the approximate 
optimal of the offline problem $\mathcal{P}_1$. \textit{SCPA} \cite{ns-game} is a state-of-the-art auction-based algorithm. We also design 
online algorithms \textit{Myopic Slicing} and \textit{Random Slicing} as baselines.
\begin{itemize}
    \item \textit{CVX} (offline \& centralized): This refers to the algorithm behinds CVXPY\footnote{\texttt{https://www.cvxpy.org/}}. We use this as a professional solver to obtain 
    the approximately optimal solution of the global offline problem $\mathcal{P}_1$.
    \item \textit{Heuristic} (offline \& centralized): We take Genetic Algorithm (GA) to obtain the approximate optimal solution of $\mathcal{P}_1$.
    \item \textit{SCPA} (offline \& decentralized)\cite{ns-game}: To adapt this algorithm to our model, we made some simple deformation. In this algorithm, 
    all the tenants and the MVNO get together. The bids are the utilities. Specifically, in each bidding around, each tenant calculate its utility. 
    If the utility is positive, it sends $x_n = 1$ and $\{d_n^c\}_{\forall c \in \mathcal{C}}$ to the MVNO. The MVNO selects the exact tenant 
    which can maximize the its own utility and accepts the transaction if resource surplus is satisfied. All the left tenants are rejected. The 
    procedure ends when no tenant has the willingness to bid.
    \item \textit{Myopic Slicing (MS)} (online \& decentralized): This algorithm is almost the same with \textsf{DPoS}, expect the pricing 
    functions. The pricing functions are designed as follows: 
    $\forall c \in \mathcal{C}, \phi'_c (y) \triangleq \frac{\underline{p_c} + \overline{p_c}}{C} y$ when $y \leq 1$, otherwise $+\infty$.
    \item \textit{Random Slicing (RS)} (online): Each time when a new tenant arrives, randomly set $x_n$ as $0$ or $1$. Note that if $x_n = 1$, 
    the resource surplus must be satisfied.
\end{itemize}
The following analyze is based on the average returns of 1000 trials. 

\subsection{Performace Verification}\label{sec5.2}
\begin{figure}[htbp]
    \centerline{\includegraphics[width=3.5in]{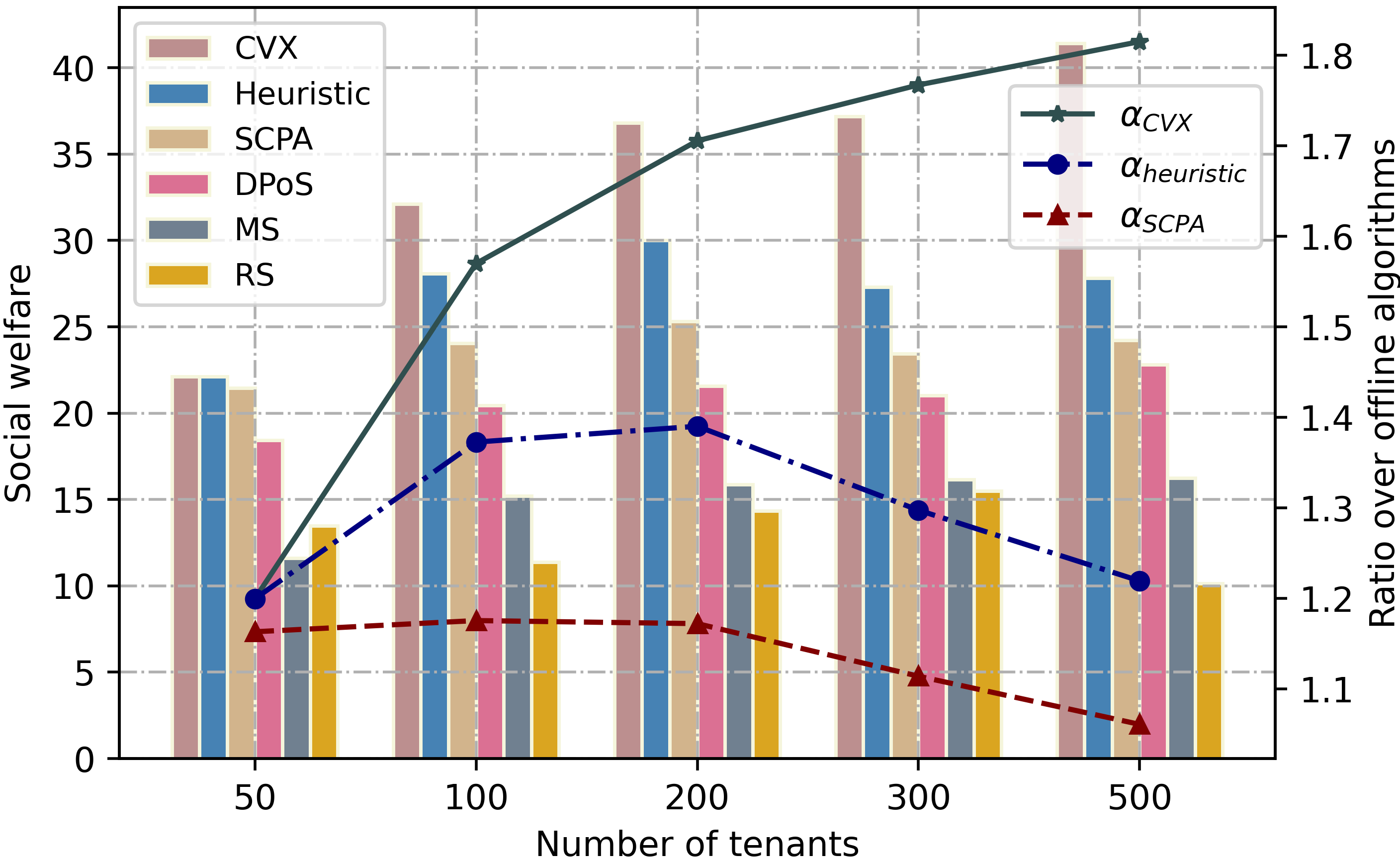}}
    \caption{The social welfare achieved by each algorithm and the ratio of social welfare achieved by each offline algorithm to \textsf{DPoS}, 
    under different number of tenants.}
    \label{fig-exp1}
\end{figure}

We firstly analyze the performance under different scales of tenants. As shown in Fig. \ref{fig-exp1}, all the offline algorithms 
outperform the online algorithms. Therewith, CVX achieves the highest social welfare whatever the number of tenants. In the following, 
we will simply take CVX as the optimal solution. It is interesting to find that both Heuristic and SCPA show a trend of performance 
decline as the number of tenants increase. For Heuristic, as the solution space grows exponentially with the increase of tenant size, 
it becomes more difficult to find the approximate optimal solution under the constraints of iteration times and population size. 
When the scale of tenants grows, the performance of all the online algorithms present a rising trend. This is becasue the transaction 
success rate increases (although not by as much) with scale under the well-designed pricing functions. Further, we can find that
\textsf{DPoS} not only achieves the highest social welfare among all the online algorithms, but also shows the \textit{close-to-offline-optimal} 
performance. Specifically, we define the indicator $\alpha_{\textrm{\textit{CVX}}}$, $\alpha_{\textrm{\textit{heuristic}}}$, and 
$\alpha_{\textrm{\textit{SCPA}}}$, where each is the ratio of the social welfare achieved by CVX, Heuristic, and SCPA to \textsf{DPoS}, 
respectively. From Fig. \ref{fig-exp1} we find that even in the worst case ($N = 500$), the gap between CVX and \textsf{DPoS} is 
only $0.815\times$. This ratio is much better (lower) compared with previous work \cite{online-ns-auction}. Compared with the popular 
offline Heuristic (GA), the gap is $0.390\times$ at the peak ($N = 200$). Compared with the state-of-the-art offline auciton-based 
algorithm SCPA \cite{ns-game}, the gap is $0.175\times$ at the peak ($N = 100$). Becasue of the performance downgrade of Heuristic and SCPA, 
the ratio $\alpha_{\textrm{\textit{heuristic}}}$ and $\alpha_{\textrm{\textit{SCPA}}}$ shows a tendency to increase first and then decrease. 

\begin{figure}[htbp]
    \centerline{\includegraphics[width=3.5in]{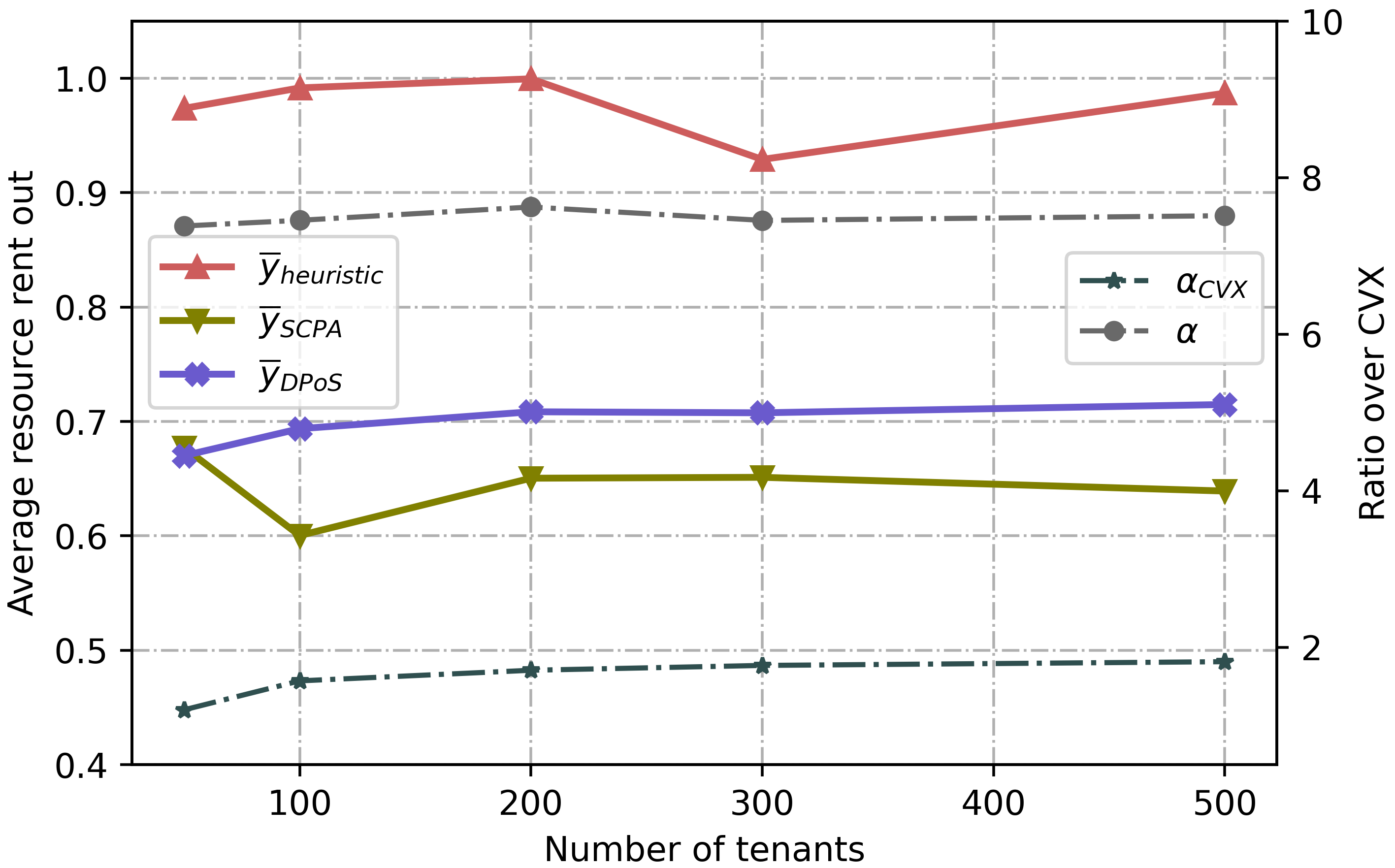}}
    \caption{Left $y$-axis: The average rental rate over 3 kinds of resources of Heuristic, SCPA, and DPoS. We do not draw the rental 
    rate of CVX because the value is close to $1$ under any circumstances.
    Right $y$-axis: the comparison of $\alpha_{\textrm{\textit{CVX}}}$ and the theoretical worst-case competitive ratio $\alpha$.}
    \label{fig-exp2}
\end{figure}

Fig. \ref{fig-exp2} demonstrates that Heuristic has a near-to-1 rental rate whatever the number of tenants but SCPA's and \textsf{DPoS}'s 
rental rate are much lower ($64.37\%$ and $69.89\%$ in average, respectively). However, from Fig. \ref{fig-exp1} we have concluded that the performance 
of Heuristic is much inferior to the optimal especially when $N$ is $500$. Thus, we can conclude that there is no \textit{linear} relationship between 
the sum of net profits and the transaction success rate. In fact, this conclusion can also be draw by observing the analytic form of 
social welfare defined in $\mathcal{P}_1$. Besides, the scale of tenants has no significant impact on the rental rate, whether 
it is an offline algorithm, or \textsf{DPoS}. Another interesting point is that under normal circumstances, the worst-case theoretical guarantee, 
i.e. the competitive ratio calculated according to \eqref{w_c} and \eqref{worst-ratio}, is far from need. 

\begin{figure}[htbp]
    \centerline{\includegraphics[width=3.5in]{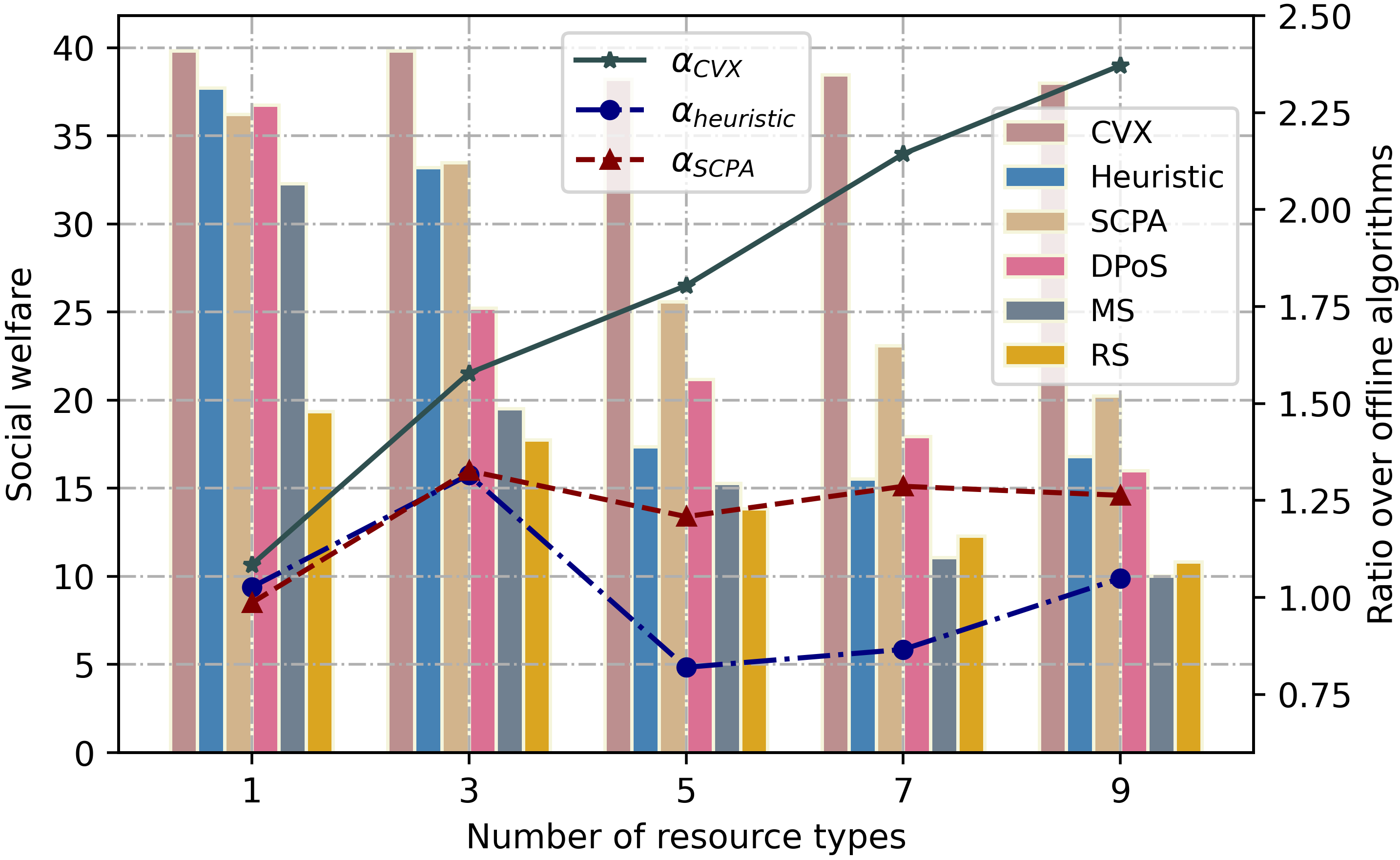}}
    \caption{The social welfare achieved by each algorithm and the ratio of social welfare achieved by each offline algorithm to \textsf{DPoS}, 
    under different number of resource types.}
    \label{fig-exp3}
\end{figure}

\begin{figure}[htbp]
    \centerline{\includegraphics[width=3.5in]{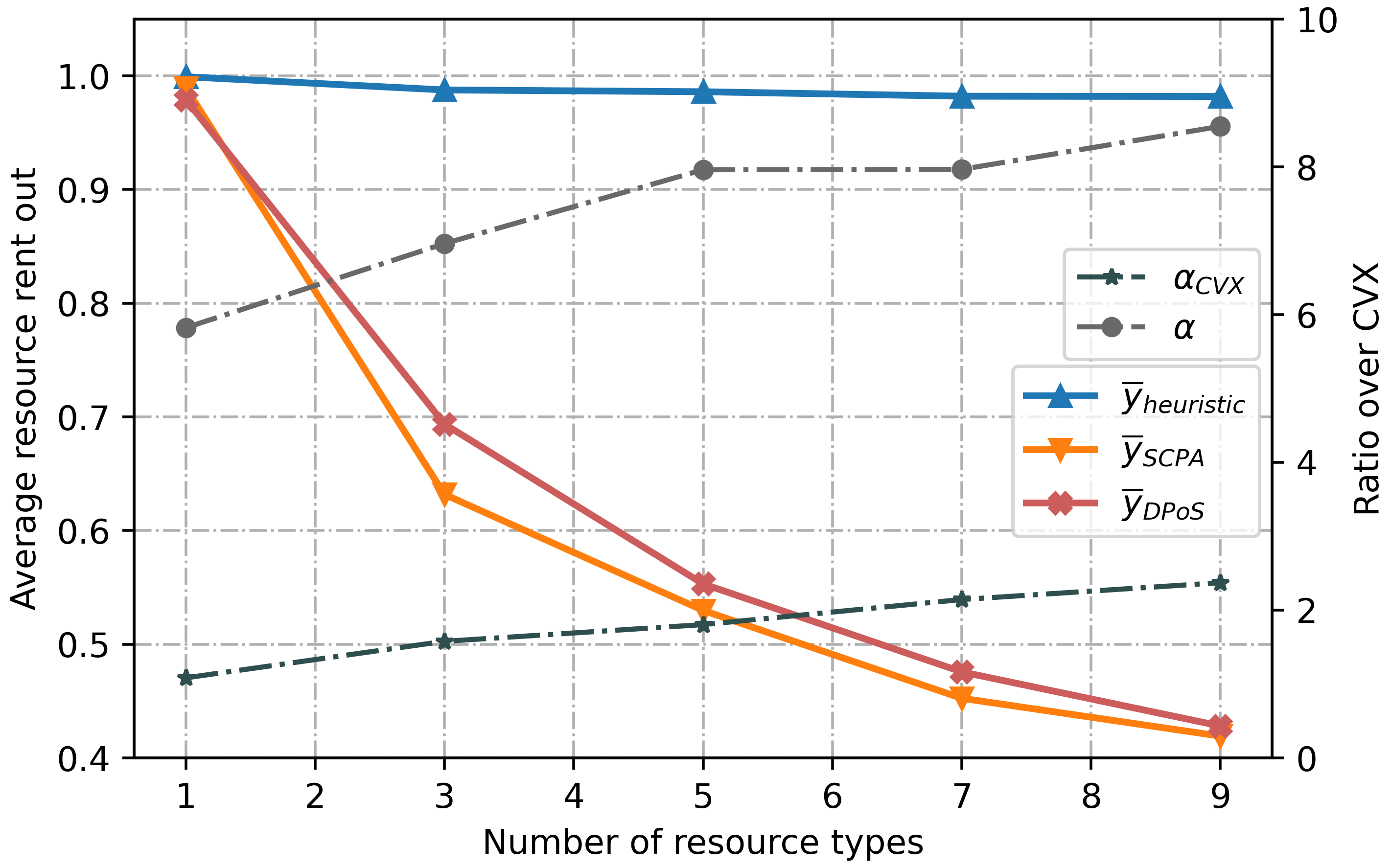}}
    \caption{Left $y$-axis: The average rental rate over 3 kinds of resources of Heuristic, SCPA, and DPoS.
    Right $y$-axis: the comparison of $\alpha_{\textrm{\textit{CVX}}}$ and the theoretical worst-case competitive ratio $\alpha$.}
    \label{fig-exp4}
\end{figure}

\begin{table*}[htbp]   
    \begin{center}
    \caption{\label{para2}Comparsion of transferred data size and algorithm's running time under default parameter settings.}
    \label{tab3}
    \begin{tabular}{c|c|c|c|c|c|c}
        \toprule
         & {\textsf{\textbf{CVX}}} & {\textsf{\textbf{Heuristic}}} & {\textsf{\textbf{SCPA}}} 
        & {\textsf{\textbf{DPoS}}} & {\textsf{\textbf{MS}}} & {\textsf{\textbf{RS}}}\\[+0.1mm]
        \midrule
        \textsf{input form} & offline & offline & offline & online & online & online\\[+0.7mm]
        \textsf{architecture} & centralized & centralized & decentralized & decentralized & decentralized & -\\[+0.7mm]
        \textsf{transferred data size} & 4.16KB & 4.16KB & 4.16KB & 1.92KB & 1.92KB & -\\[+0.7mm]
        \textsf{running time} & 78.81 & 2172.35 & 24.43 & 1 & 0.93 & 0.48\\[+0.7mm]
        $\alpha_{\textrm{\textit{CVX}}}$ & 1 & 1.199 & 1.189 & 1.578 & 2.04 & 2.47\\[+0.7mm]
        \bottomrule   
    \end{tabular}  
    \end{center}
\end{table*}

In the following we analyze the performance of \textsf{DPoS} under different scale of resource types $C$. From Fig. \ref{fig-exp3}, firstly, 
we find that \textsf{DPoS} is still the best online algorithm and has a close performance to Heuristic and SCPA. When $C = 1$, \textsf{DPoS} can 
achieve near-to-offline-optimal performance! Secondly, all the algorithms show a downward trend when the number of resource types increase, 
except CVX. This is becasue each tenant has requirements on all the resource types, and the increase in resource types significantly 
reduces the probability of requirements being satisfied. Ulteriorly, the transaction success rate reduces significantly. The phenomena 
can also be found in Fig. \ref{fig-exp4}. For online scenarios, the phenomena is amplified by the randomness of arrival sequence of tenants. 
Thus, online algorithms perform more unsatisfied. Even though, the advantage of \textsf{DPoS} is clear. In the worst case, i.e., when 
$C = 9$, the ratio $\alpha_{\textrm{\textit{CVX}}}$ is $2.37$, which is still acceptable for online algorithms. It even outperforms 
the offline algorithm Heuristic when $C$ is $5$ and $7$ by $18.00\%$ and $13.40\%$, respectively.

\begin{figure}[htbp]
    \centerline{\includegraphics[width=2.2in]{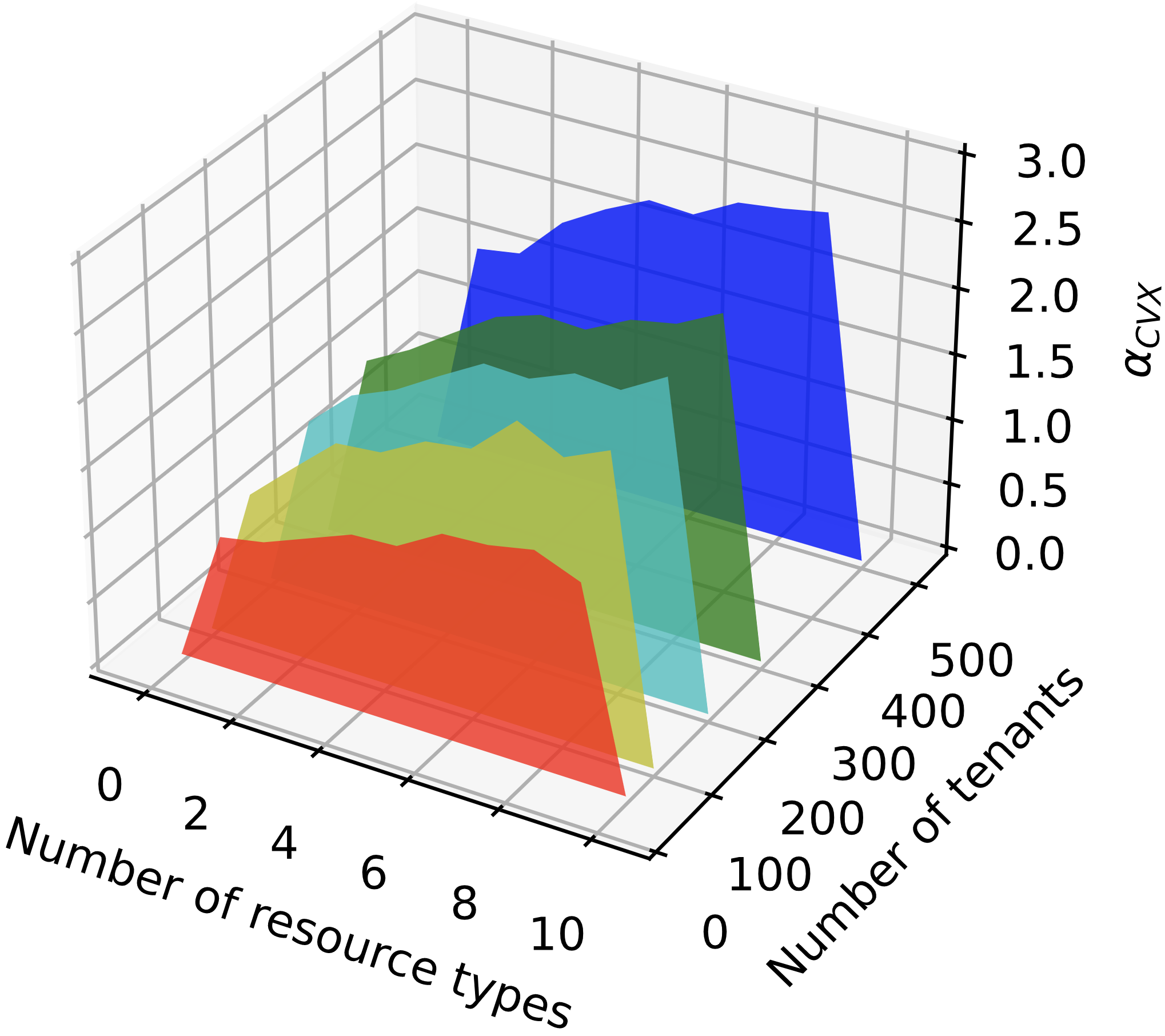}}
    \caption{The ratio of social welfare achieved by \textsf{DPoS} to the optimal, CVX, under different scales of tenants and resource types.}
    \label{fig-exp5}
\end{figure}

\begin{figure}[htbp]
    \centerline{\includegraphics[width=3.2in]{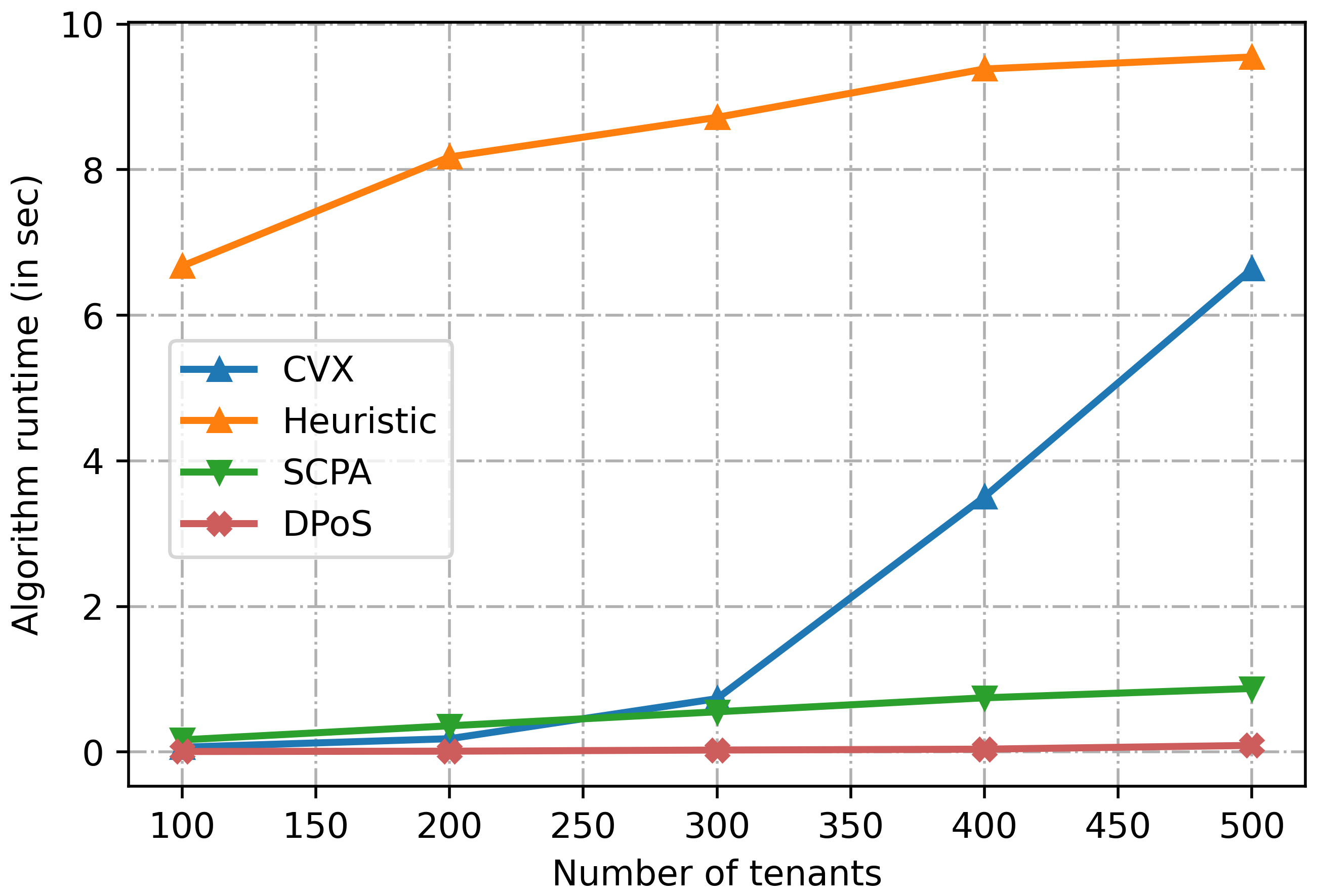}}
    \caption{The runtime of each algorithm under different number of tenants.}
    \label{fig-add1}
\end{figure}

\begin{figure}[htbp]
    \centerline{\includegraphics[width=3.2in]{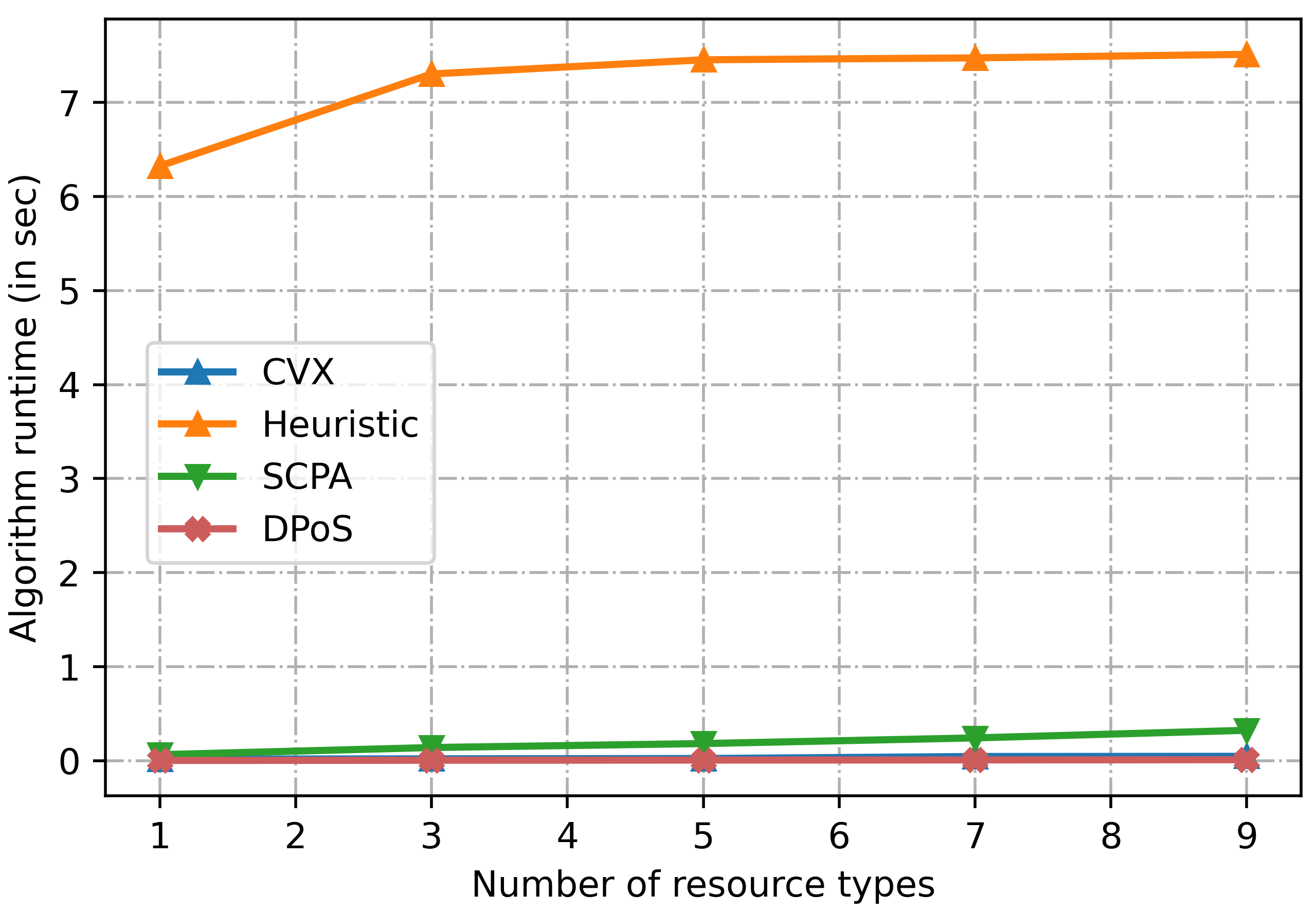}}
    \caption{The runtime of each algorithm under different number of resource types.}
    \label{fig-add2}
\end{figure}

Fig. \ref{fig-exp5} demonstrates the impact of scales of tenants and resource types on the performance of \textsf{DPoS} comprehensively. 
In general, the gap between \textsf{DPoS} and the offline optimal increases with the increasing scale of the problem. When $C$ is $1$ and $N$ is $50$, 
what \textsf{DPoS} achieves is exactly the offline optimal. When $C$ is $9$ and $N$ is $500$, the gap is the highest, which reachs $1.57\times$.
Further, we can find that the ratio grows faster with resource types than with tenant size. We leave the design of resource type-scalable 
pricing functions as future work. Table \ref{tab3} comapares all the algorithms from multiple angles, including social welfare achieved, 
cross-agent communication data size, and algorithm running time. The amount of data transferred by the decentralized online algorithm refers 
to the amount of data communicated between tenants and the MVNO. Meanwhile, the amount of data transferred by the centralized algorithm is 
all data related to problem $\mathcal{P}_1$. The data size is calculated as 4 bytes for each value. Note that we normalize the running time 
of \textsf{DPoS} to $1$. We can find that the superiority of CVX and Heuristic are based on a lot of computing time overhead. By contrast, 
\textsf{DPoS} achieves a satisfactory balance between the overheads the performance. 
In addition to the 4-th line of Table \ref{tab3}, Fig. \ref{fig-add1} and Fig. \ref{fig-add2} also verify the linear algorithmic 
runtime of \textsf{DPoS} intuitively.

\subsection{Sensitivity Analysis}\label{sec5.3}
In this subsection, we analyze the sensitivity of \textsf{DPoS} under different environment parameters settings. 

\begin{figure}[htbp]
    \centerline{\includegraphics[width=3.5in]{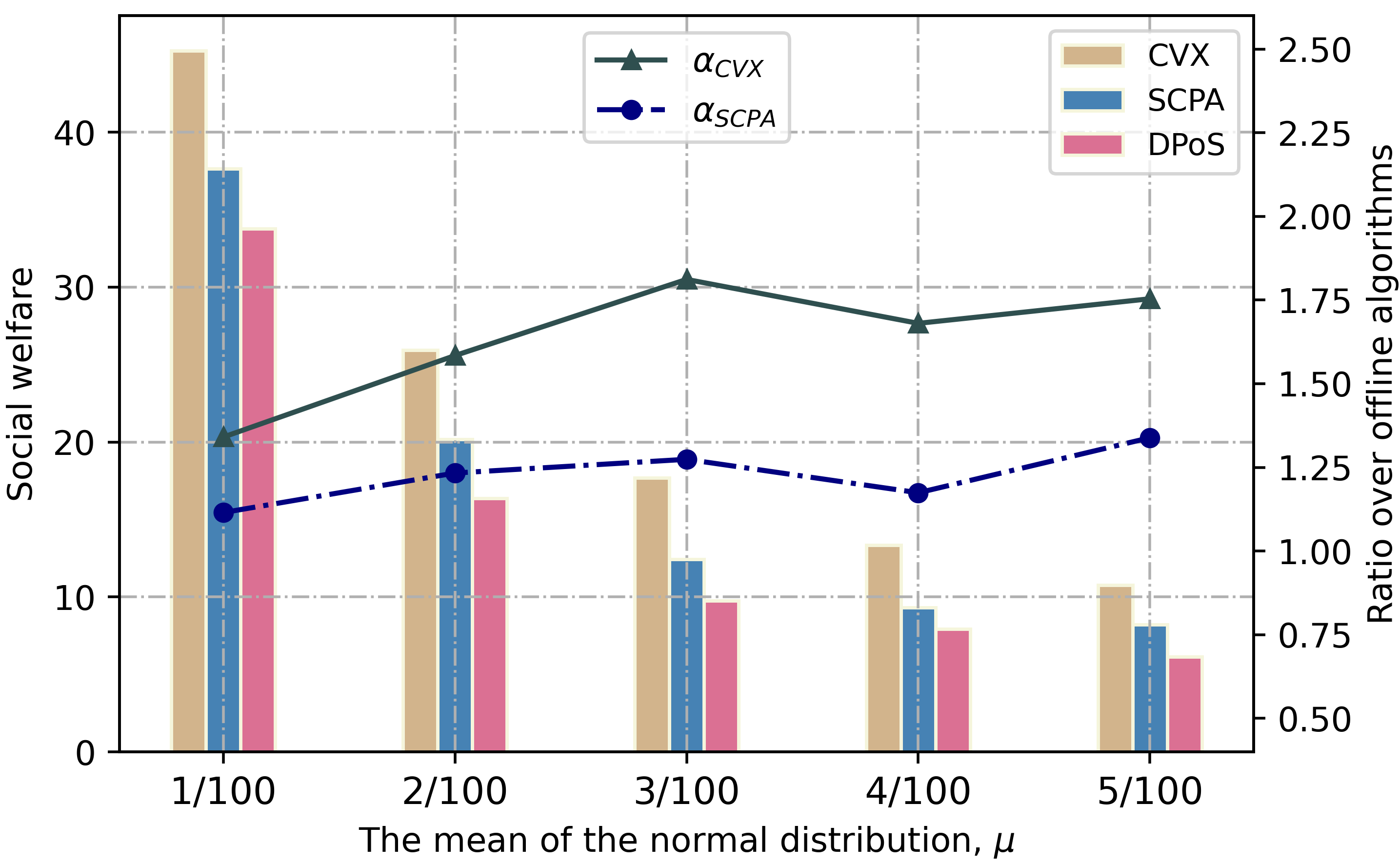}}
    \caption{The social welfare achieved by each algorithm and the ratio of social welfare achieved by CVX and SCPA to \textsf{DPoS}, 
    under different sampling of $\{d_n^c\}_{\forall c \in \mathcal{C}}$.}
    \label{fig-exp6}
\end{figure}

Fig. \ref{fig-exp6} demonstrates the impact of tenants' resource requirements. The $x$-axis is the mean value $\mu$ of the Normal distribution 
$N(\mu, \sigma=\frac{1}{N^2})$ where $N$ is $100$. We find that when the resource requirements increase, the transaction success rate 
decreases, which further decreases the social welfare achieved. It is interesting that the social welfare achieved by CVX also decreases 
significantly when tenants' resource requirements increase. This phenomenon indicates that the competition among tenants for resources 
significantly reduces the feasible solution space. Even so, the ratio on social welfare is stable no matter how the resource requirements change.

\begin{figure}[htbp]
    \centerline{\includegraphics[width=3.5in]{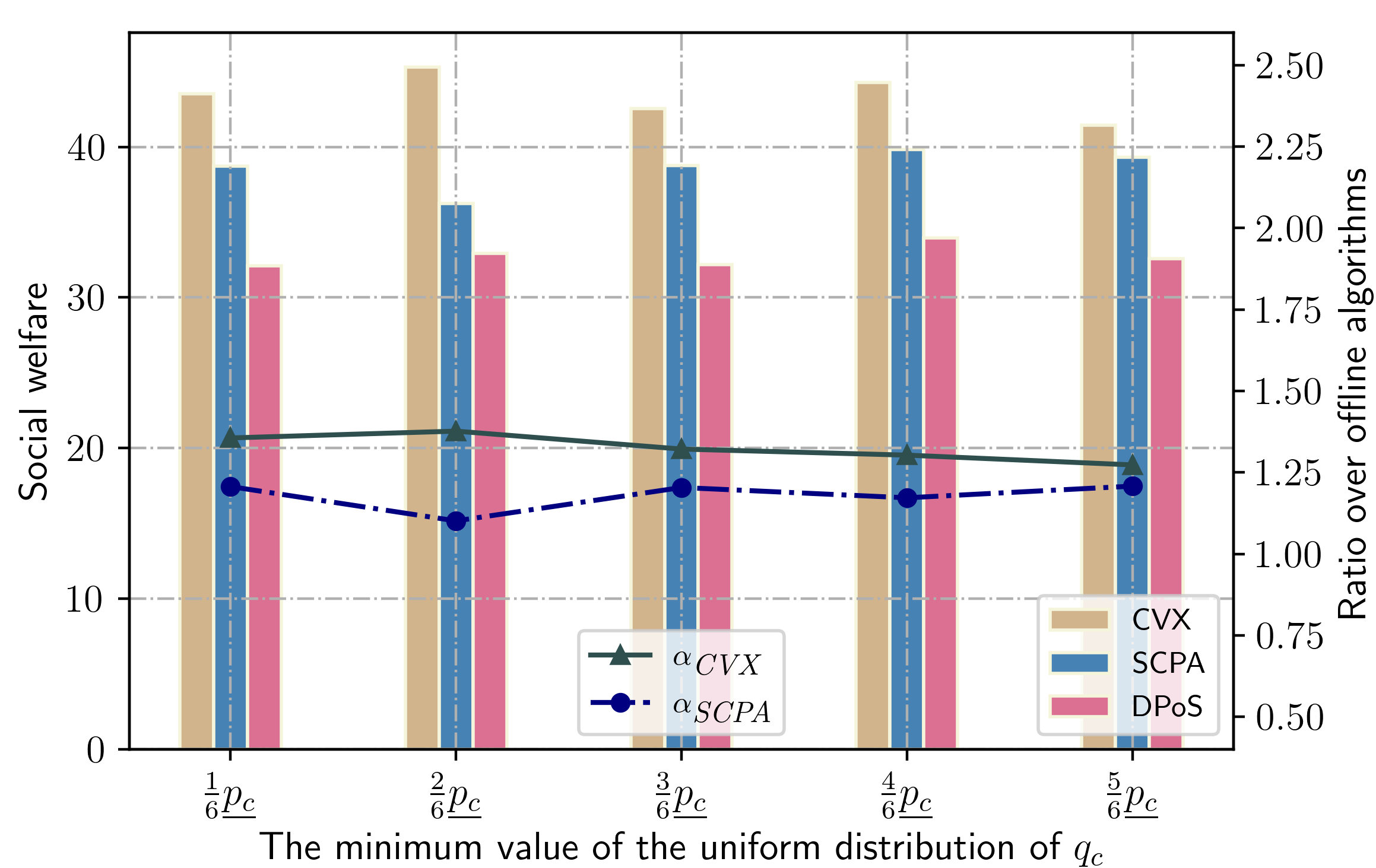}}
    \caption{The social welfare achieved by each algorithm and the ratio of social welfare achieved by CVX and SCPA to \textsf{DPoS}, 
    under different sampling of $\{q_c\}_{\forall c \in \mathcal{C}}$.}
    \label{fig-exp7}
\end{figure}

\begin{figure}[htbp]
    \centerline{\includegraphics[width=3.5in]{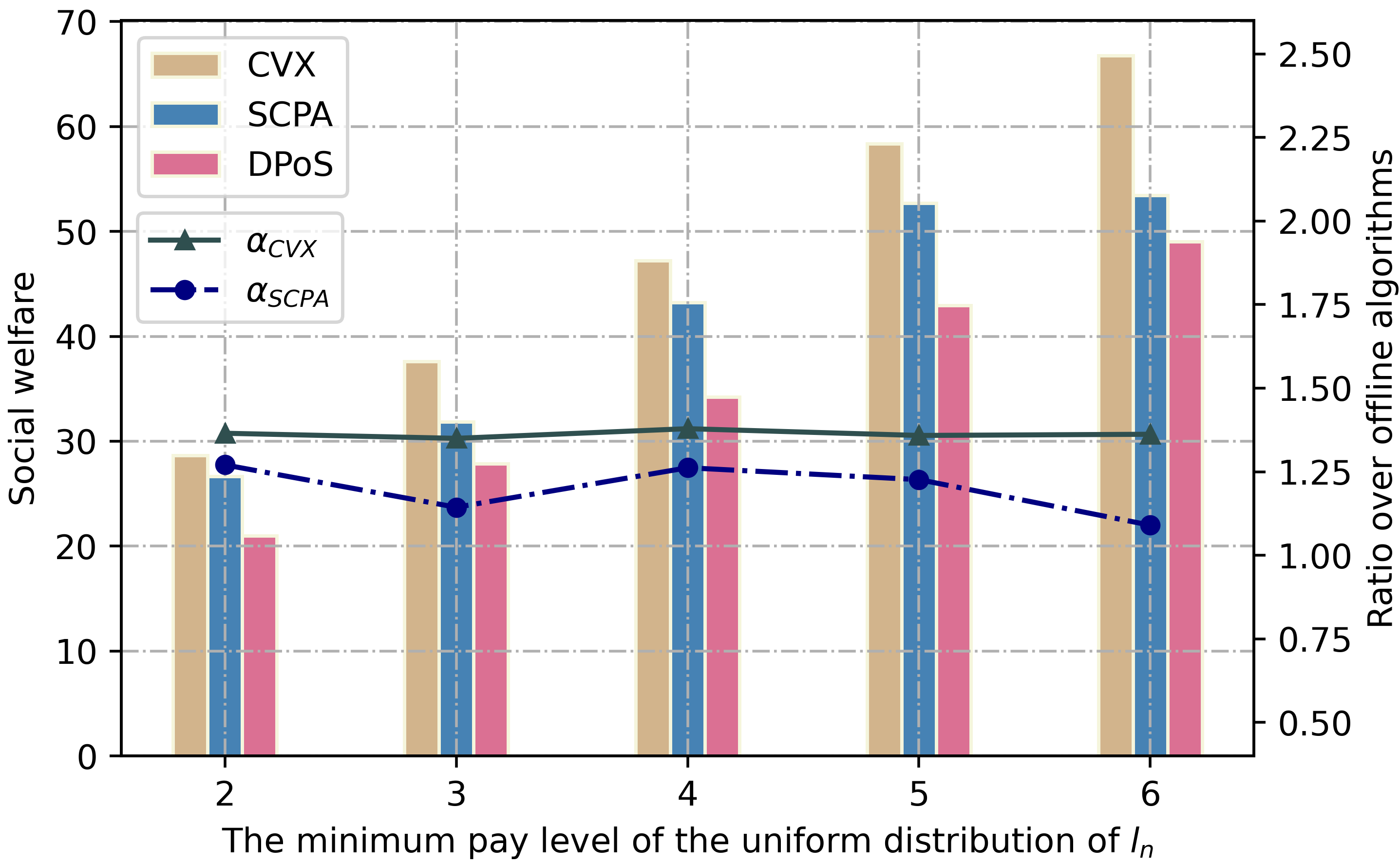}}
    \caption{The social welfare achieved by each algorithm and the ratio of social welfare achieved by CVX and SCPA to \textsf{DPoS}, 
    under different sampling of pay levels $\{l_n\}_{\forall n \in \mathcal{N}}$.}
    \label{fig-exp8}
\end{figure}

Fig. \ref{fig-exp7} and Fig. \ref{fig-exp8} demonstrate the impact of $\{q_c\}_{\forall c \in \mathcal{C}}$ and $\{l_n\}_{\forall n \in \mathcal{N}}$. 
We can find that the ratio on social welfare has a smooth variation. Considering that their impacts are minor, no more detailed discussion will 
be launched.

All the experiment results in this subsection show the robustness of \textsf{DPoS}.

\section{Related Works}\label{sec6}
Network slicing is widely accepted as an architectural enabling technology for 5G by industry and standardization communities 
\cite{5G-archi}\cite{ns-survey2}\cite{ns-survey3}\cite{ns-survey}. The idea is to \textit{slice} the physical resources of the mobile 
networks into logical network functions, and orchestrate them to support diversified over-the-top services. 
Previous works on network slicing mainly focus on the architectural aspects, while efficient resource allocation and sharing, 
which has been identified as a key issue by the Next Generation Mobile Network (NGMN) alliance \cite{5G-archi2}, lags behind. 

A number of studies have emerged in recent years to fill the gap, especially for mobile network slicing \cite{ran-slice}\cite{edge-slice2}
\cite{edge-slice} and core network slicing \cite{core-slice}\cite{core-add1}\cite{core-add3}. Overall, these works formulate 
a non-convex combinatorial problem to maximize the utilities of involved business players. Take \cite{ran-slice} as an example, 
the authors defined the utility according to the satisfaction of multiple slice resource demands (SRDs). They formulated the resource 
sharing problem as a Mixed Integer Linear Programming (MILP) and proposed a two-step approach (provisioning-and-deployment) to solve 
it efficiently. Similarly, Caballero et al. proposed a dynamic resource allocation algorithm based on the weighted proportionally fairness, 
also for the RAN resources \cite{edge-slice2}. Based on this algorithm, they devised a practical approach with limited computational 
information and handoff overheads. Further, the authors verified the approximate optimality of the approach with both theoretical proof and 
extensive simulations. In addition to the heuristics designed by the above mentioned works, AI-based optimization has been gaining in popularity. 
For example, Yan et al. resorted to deep reinforcement learning (DRL) to formulate an intelligent resource scheduling strategy, iRSS, for 5G RAN 
slicing \cite{DL}. They take deep neural networks to perform large time-scale resource allocation while the reinforce 
agent performs online resource scheduling to predict network states and dynamics. Likewise, the authors of \cite{resource-manage-DQN} 
also designed a DRL-based algorithm, to perform corss-slice resource sharing.

In addition to the centralized and fine-tuned algorithms, a substantial literature designed the network slicing algorithms based on 
economic frameworks, especially the auction-related mechanisms \cite{ns-business}\cite{nfv} \cite{nfv-offline}\cite{congest-game-slice}
\cite{ns-game}\cite{auction}. These algorithms are usually decentralized, easy-to-use and simply constructed. In these works, the tenants 
sequentially compete and bid for the network resources. The utilization of auction mechanism usually integrate tightly with dynamic pricing 
and game model \cite{auction-or-posted}. For example, Wang et al. solved the joint efficiency and revenue maximization problem with a 
varying-pricing policy \cite{s1}. They designed a decentralized algorithm, run by each player, to maximize the net social welfare. In \cite{ns-game}, 
the authors designed a non-cooperative game where each tenant reacts to the user allocations of the other tenants so as to maximize 
its own utility selfishly. Existing works mainly resort to Fisher market \cite{fisher2}, where strategic players anticipate the impact 
of their bids. Besides, VCG-Kelly mechanisms and their derivatives \cite{vcg-kelly} are also popular for slice resource allocation and 
sharing \cite{ns-game}\cite{online-ns-auction}. In Kelly's mechanism, the bidders bid for prices, and the resources are allocated to them 
according to their bids. In VCG mechanism, in a different way, the bids are the utility of involved players. We find that existing auction-based 
works are mainly designed for offline markets, where all the tenants participate the auction and bid for their interests sequentially. 
Even so, we still discover an online auction-based resource allocation algorithm, proposed in \cite{online-ns-auction}. The authors model 
the slicing resource allocation problem as an online winner determination problem, with aim to maximize the social welfare of auction 
participants. However, what the authors of \cite{online-ns-auction} proposed is a centralized algorithm, where the bidding and 
privacy-relevant information has to be collected by the MVNO.

Our work is based on the posted price mechanism \cite{posted-mechanism}, under the principle of \textit{take-it-or-leave-it}. 
Compared with fined-tuned heuristics and DRL-based works, our algorithm has fairly low complexity and is well-suited for online network slicing 
scenarios. Besides, the time-consuming repeat bidding between tenants and the MVNO is not required compared with auction-based works. 
In addition, our algorithm provides each business player an agent, which can be deployed in a realistic online market directly without any modification.

\section{Concluding Remarks}\label{sec7}
We presented a decentralized and low-complexity online slicing algorithm, \textsf{DPoS}, by virtue of the primal-dual approach and posted price 
mechanism. Our goal was to address the problem of the high complexity, privacy leakage, and unrealistic offline setting of current network slicing algorithms.
We firstly presented the global offline social welfare maximization problem. Then, we relax the original combinatorial problem to a convex primal problem 
and give its dual. Based on the alternative update of primal and dual variables, \textsf{DPoS} maximizes the social welfare with a 
$O \big( \max_{c \in \mathcal{C}} \{\ln \sum_{c' \in \mathcal{C}} (\overline{p_{c'}} - q_{c'}) - \ln (\underline{p_c} - q_c) \} \big)$ gap 
in worst case. By giving back the decision-making power to each player, \textsf{DPoS} stops the privacy leakage from the source. This decentralized 
property also erases the heavy burden to solve a centralized offline optimization algorithm, which is often of high complexity. In addition to 
the efficiency, the competitive ratio of \textsf{DPoS} is the optimal over all the online algorithms. The extensive simulation further verify that 
\textsf{DPoS} can not only achieve close-to-offline-optimal performance, but also have much lower algorithmic overheads compared with contrast algorithms.

\ifCLASSOPTIONcompsoc
  \section*{Acknowledgments}
\else
  \section*{Acknowledgment}
\fi
This work was partially supported by the National Science Foundation of China (No. U20A20173 and No. 61772461), the National Key 
Research and Development Program of China (No. 2019YFD1101105) and Natural Science Foundation of Zhejiang Province (No. LR18F020003).
Schahram Dustdar's work is supported by the Zhejiang University Deqing Institute of Advanced technology and Industrilization (ZDATI). 

\bibliographystyle{IEEEtran}
\bibliography{IEEEabrv,ref.bib}

\begin{thebibliography}{10}
\providecommand{\url}[1]{#1}
\csname url@samestyle\endcsname
\providecommand{\newblock}{\relax}
\providecommand{\bibinfo}[2]{#2}
\providecommand{\BIBentrySTDinterwordspacing}{\spaceskip=0pt\relax}
\providecommand{\BIBentryALTinterwordstretchfactor}{4}
\providecommand{\BIBentryALTinterwordspacing}{\spaceskip=\fontdimen2\font plus
\BIBentryALTinterwordstretchfactor\fontdimen3\font minus
  \fontdimen4\font\relax}
\providecommand{\BIBforeignlanguage}[2]{{%
\expandafter\ifx\csname l@#1\endcsname\relax
\typeout{** WARNING: IEEEtran.bst: No hyphenation pattern has been}%
\typeout{** loaded for the language `#1'. Using the pattern for}%
\typeout{** the default language instead.}%
\else
\language=\csname l@#1\endcsname
\fi
#2}}
\providecommand{\BIBdecl}{\relax}
\BIBdecl

\bibitem{5G-archi}
{5G PPP Architecture Working Group}, ``View on 5g architecture: Version 3.0,''
  \url{https://doi.org/10.5281/zenodo.3265031}, Feb 2020.

\bibitem{ns-survey2}
M.~{Richart}, J.~{Baliosian}, J.~{Serrat}, and J.~{Gorricho}, ``Resource
  slicing in virtual wireless networks: A survey,'' \emph{IEEE Transactions on
  Network and Service Management}, vol.~13, no.~3, pp. 462--476, 2016.

\bibitem{ns-survey3}
X.~{Foukas}, G.~{Patounas}, A.~{Elmokashfi}, and M.~K. {Marina}, ``Network
  slicing in 5g: Survey and challenges,'' \emph{IEEE Communications Magazine},
  vol.~55, no.~5, pp. 94--100, 2017.

\bibitem{ns-survey}
I.~{Afolabi}, T.~{Taleb}, K.~{Samdanis}, A.~{Ksentini}, and H.~{Flinck},
  ``Network slicing and softwarization: A survey on principles, enabling
  technologies, and solutions,'' \emph{IEEE Communications Surveys Tutorials},
  vol.~20, no.~3, pp. 2429--2453, 2018.

\bibitem{ns-business}
D.~{Bega}, M.~{Gramaglia}, A.~{Banchs}, V.~{Sciancalepore}, K.~{Samdanis}, and
  X.~{Costa-Perez}, ``Optimising 5g infrastructure markets: The business of
  network slicing,'' in \emph{IEEE INFOCOM 2017 - IEEE Conference on Computer
  Communications}, 2017, pp. 1--9.

\bibitem{mobicom-slicing-paper}
C.~Marquez, M.~Gramaglia, M.~Fiore, A.~Banchs, and X.~Costa-Perez, ``How should
  i slice my network?: A multi-service empirical evaluation of resource sharing
  efficiency,'' in \emph{Proceedings of the 24th Annual International
  Conference on Mobile Computing and Networking}, ser. MobiCom '18, New York,
  NY, USA, 2018, pp. 191--206.

\bibitem{nfv}
X.~{Zhang}, Z.~{Huang}, C.~{Wu}, Z.~{Li}, and F.~C.~M. {Lau}, ``Online
  stochastic buy-sell mechanism for vnf chains in the nfv market,'' \emph{IEEE
  Journal on Selected Areas in Communications}, vol.~35, no.~2, pp. 392--406,
  2017.

\bibitem{nfv-offline}
S.~{Gu}, Z.~{Li}, C.~{Wu}, and C.~{Huang}, ``An efficient auction mechanism for
  service chains in the nfv market,'' in \emph{IEEE INFOCOM 2016 - The 35th
  Annual IEEE International Conference on Computer Communications}, 2016, pp.
  1--9.

\bibitem{ns-algorithm-pers}
S.~{Vassilaras}, L.~{Gkatzikis}, N.~{Liakopoulos}, I.~N. {Stiakogiannakis},
  M.~{Qi}, L.~{Shi}, L.~{Liu}, M.~{Debbah}, and G.~S. {Paschos}, ``The
  algorithmic aspects of network slicing,'' \emph{IEEE Communications
  Magazine}, vol.~55, no.~8, pp. 112--119, 2017.

\bibitem{ran-slice}
Q.~{Luu}, S.~{Kerboeuf}, A.~{Mouradian}, and M.~{Kieffer}, ``A coverage-aware
  resource provisioning method for network slicing,'' \emph{IEEE/ACM
  Transactions on Networking}, pp. 1--14, 2020.

\bibitem{edge-slice2}
P.~{Caballero}, A.~{Banchs}, G.~{de Veciana}, and X.~{Costa-Pérez},
  ``Multi-tenant radio access network slicing: Statistical multiplexing of
  spatial loads,'' \emph{IEEE/ACM Transactions on Networking}, vol.~25, no.~5,
  pp. 3044--3058, 2017.

\bibitem{tsc-edge}
H.~{Zhao}, S.~{Deng}, Z.~{Liu}, J.~{Yin}, and S.~{Dustdar}, ``Distributed
  redundancy scheduling for microservice-based applications at the edge,''
  \emph{IEEE Transactions on Services Computing}, pp. 1--1, 2020.

\bibitem{tmc-edge}
S.~{Deng}, Z.~{Xiang}, J.~{Taheri}, K.~A. {Mohammad}, J.~{Yin}, A.~{Zomaya},
  and S.~{Dustdar}, ``Optimal application deployment in resource constrained
  distributed edges,'' \emph{IEEE Transactions on Mobile Computing}, pp. 1--1,
  2020.

\bibitem{core-slice}
D.~A. {Chekired}, M.~A. {Togou}, L.~{Khoukhi}, and A.~{Ksentini},
  ``5g-slicing-enabled scalable sdn core network: Toward an ultra-low latency
  of autonomous driving service,'' \emph{IEEE Journal on Selected Areas in
  Communications}, vol.~37, no.~8, pp. 1769--1782, 2019.

\bibitem{core-add1}
M.~R. {Sama}, X.~{An}, Q.~{Wei}, and S.~{Beker}, ``Reshaping the mobile core
  network via function decomposition and network slicing for the 5g era,'' in
  \emph{2016 IEEE Wireless Communications and Networking Conference}, 2016, pp.
  1--7.

\bibitem{core-add3}
D.~{Sattar} and A.~{Matrawy}, ``Optimal slice allocation in 5g core networks,''
  \emph{IEEE Networking Letters}, vol.~1, no.~2, pp. 48--51, 2019.

\bibitem{edge-slice}
P.~L. {Vo}, M.~N.~H. {Nguyen}, T.~A. {Le}, and N.~H. {Tran}, ``Slicing the
  edge: Resource allocation for ran network slicing,'' \emph{IEEE Wireless
  Communications Letters}, vol.~7, no.~6, pp. 970--973, 2018.

\bibitem{resource-manage-DQN}
X.~{Chen}, Z.~{Zhao}, C.~{Wu}, M.~{Bennis}, H.~{Liu}, Y.~{Ji}, and H.~{Zhang},
  ``Multi-tenant cross-slice resource orchestration: A deep reinforcement
  learning approach,'' \emph{IEEE Journal on Selected Areas in Communications},
  vol.~37, no.~10, pp. 2377--2392, Oct 2019.

\bibitem{tpds-burst}
S.~{Deng}, C.~{Zhang}, C.~{Li}, J.~{Yin}, S.~{Dustdar}, and A.~Y. {Zomaya},
  ``Burst load evacuation based on dispatching and scheduling in distributed
  edge networks,'' \emph{IEEE Transactions on Parallel and Distributed
  Systems}, vol.~32, no.~8, pp. 1918--1932, 2021.

\bibitem{heuristic}
B.~{Han}, J.~{Lianghai}, and H.~D. {Schotten}, ``Slice as an evolutionary
  service: Genetic optimization for inter-slice resource management in 5g
  networks,'' \emph{IEEE Access}, vol.~6, pp. 33\,137--33\,147, 2018.

\bibitem{DL}
M.~{Yan}, G.~{Feng}, J.~{Zhou}, Y.~{Sun}, and Y.~{Liang}, ``Intelligent
  resource scheduling for 5g radio access network slicing,'' \emph{IEEE
  Transactions on Vehicular Technology}, vol.~68, no.~8, pp. 7691--7703, 2019.

\bibitem{add3}
X.-L. Huang, X.~Ma, and F.~Hu, ``Machine learning and intelligent
  communications,'' \emph{Mobile Networks and Applications}, vol.~23, no.~1,
  pp. 68--70, 2018.

\bibitem{edge-intelligence}
S.~{Deng}, H.~{Zhao}, W.~{Fang}, J.~{Yin}, S.~{Dustdar}, and A.~Y. {Zomaya},
  ``Edge intelligence: The confluence of edge computing and artificial
  intelligence,'' \emph{IEEE Internet of Things Journal}, vol.~7, no.~8, pp.
  7457--7469, 2020.

\bibitem{add1}
\BIBentryALTinterwordspacing
J.~Sakuma, S.~Kobayashi, and R.~N. Wright, ``Privacy-preserving reinforcement
  learning,'' in \emph{Proceedings of the 25th International Conference on
  Machine Learning}, ser. ICML '08.\hskip 1em plus 0.5em minus 0.4em\relax New
  York, NY, USA: Association for Computing Machinery, 2008, p. 864–871.
  [Online]. Available: \url{https://doi.org/10.1145/1390156.1390265}
\BIBentrySTDinterwordspacing

\bibitem{add2}
X.~{Liu}, R.~H. {Deng}, K.~K. {Raymond Choo}, and Y.~{Yang},
  ``Privacy-preserving reinforcement learning design for patient-centric
  dynamic treatment regimes,'' \emph{IEEE Transactions on Emerging Topics in
  Computing}, vol.~9, no.~1, pp. 456--470, 2021.

\bibitem{fisher1}
\BIBentryALTinterwordspacing
Y.~K. Cheung, R.~Cole, and Y.~Tao, ``Dynamics of distributed updating in fisher
  markets,'' in \emph{Proceedings of the 2018 ACM Conference on Economics and
  Computation}, ser. EC '18.\hskip 1em plus 0.5em minus 0.4em\relax New York,
  NY, USA: Association for Computing Machinery, 2018, p. 351–368. [Online].
  Available: \url{https://doi.org/10.1145/3219166.3219189}
\BIBentrySTDinterwordspacing

\bibitem{fisher2}
R.~Cole, N.~Devanur, V.~Gkatzelis, K.~Jain, T.~Mai, V.~V. Vazirani, and
  S.~Yazdanbod, ``Convex program duality, fisher markets, and nash social
  welfare,'' in \emph{Proceedings of the 2017 ACM Conference on Economics and
  Computation}, 2017, pp. 459--460.

\bibitem{VCG}
A.~Mu'Alem and N.~Nisan, ``Truthful approximation mechanisms for restricted
  combinatorial auctions,'' \emph{Games and Economic Behavior}, vol.~64, no.~2,
  pp. 612--631, 2008.

\bibitem{vcg-kelly}
S.~Yang and B.~Hajek, ``Vcg-kelly mechanisms for allocation of divisible goods:
  Adapting vcg mechanisms to one-dimensional signals,'' \emph{IEEE Journal on
  Selected Areas in Communications}, vol.~25, no.~6, pp. 1237--1243, 2007.

\bibitem{s1}
G.~{Wang}, G.~{Feng}, W.~{Tan}, S.~{Qin}, R.~{Wen}, and S.~{Sun}, ``Resource
  allocation for network slices in 5g with network resource pricing,'' in
  \emph{GLOBECOM 2017 - 2017 IEEE Global Communications Conference}, 2017, pp.
  1--6.

\bibitem{s3}
M.~{Jiang}, M.~{Condoluci}, and T.~{Mahmoodi}, ``Network slicing in 5g: An
  auction-based model,'' in \emph{2017 IEEE International Conference on
  Communications (ICC)}, 2017, pp. 1--6.

\bibitem{zhang2011game}
Y.~Zhang and M.~Guizani, \emph{Game theory for wireless communications and
  networking}.\hskip 1em plus 0.5em minus 0.4em\relax CRC press, 2011.

\bibitem{congest-game-slice}
S.~D’Oro, F.~Restuccia, T.~Melodia, and S.~Palazzo, ``Low-complexity
  distributed radio access network slicing: Algorithms and experimental
  results,'' \emph{IEEE/ACM Trans. Netw.}, vol.~26, no.~6, p. 2815–2828, Dec.
  2018.

\bibitem{nst}
{GSM Association}, ``Official document ng.116 - generic network slice template
  v4.0,''
  \url{https://www.gsma.com/newsroom/wp-content/uploads//NG.116-v4.0-2.pdf},
  Nov 2020.

\bibitem{auction-or-posted}
L.~Einav, C.~Farronato, J.~Levin, and N.~Sundaresan, ``Auctions versus posted
  prices in online markets,'' \emph{Journal of Political Economy}, vol. 126,
  no.~1, pp. 178--215, 2018.

\bibitem{posted-mechanism}
\BIBentryALTinterwordspacing
J.~Correa, P.~Foncea, R.~Hoeksma, T.~Oosterwijk, and T.~Vredeveld, ``Posted
  price mechanisms for a random stream of customers,'' in \emph{Proceedings of
  the 2017 ACM Conference on Economics and Computation}, ser. EC '17.\hskip 1em
  plus 0.5em minus 0.4em\relax New York, NY, USA: Association for Computing
  Machinery, 2017, p. 169–186. [Online]. Available:
  \url{https://doi.org/10.1145/3033274.3085137}
\BIBentrySTDinterwordspacing

\bibitem{primal-dual2}
\BIBentryALTinterwordspacing
N.~Buchbinder and J.~S. Naor, ``The design of competitive online algorithms via
  a primal–dual approach,'' \emph{Foundations and Trends® in Theoretical
  Computer Science}, vol.~3, no. 2–3, pp. 93--263, 2009. [Online]. Available:
  \url{http://dx.doi.org/10.1561/0400000024}
\BIBentrySTDinterwordspacing

\bibitem{online-mechanism}
\BIBentryALTinterwordspacing
X.~Tan, B.~Sun, A.~Leon-Garcia, Y.~Wu, and D.~H. Tsang, ``Mechanism design for
  online resource allocation: A unified approach,'' in \emph{Abstracts of the
  2020 SIGMETRICS/Performance Joint International Conference on Measurement and
  Modeling of Computer Systems}, ser. SIGMETRICS '20.\hskip 1em plus 0.5em
  minus 0.4em\relax New York, NY, USA: Association for Computing Machinery,
  2020, p. 11–12. [Online]. Available:
  \url{https://doi.org/10.1145/3393691.3394201}
\BIBentrySTDinterwordspacing

\bibitem{primal-dual1}
\BIBentryALTinterwordspacing
Z.~Huang and A.~Kim, ``Welfare maximization with production costs: A primal
  dual approach,'' \emph{Games and Economic Behavior}, vol. 118, pp. 648 --
  667, 2019. [Online]. Available:
  \url{http://www.sciencedirect.com/science/article/pii/S0899825618300344}
\BIBentrySTDinterwordspacing

\bibitem{resource-allocation-infocom17}
M.~{Leconte}, G.~S. {Paschos}, P.~{Mertikopoulos}, and U.~C. {Kozat}, ``A
  resource allocation framework for network slicing,'' in \emph{IEEE INFOCOM
  2018 - IEEE Conference on Computer Communications}, 2018, pp. 2177--2185.

\bibitem{ns-game}
P.~{Caballero}, A.~{Banchs}, G.~{De Veciana}, and X.~{Costa-Pérez}, ``Network
  slicing games: Enabling customization in multi-tenant mobile networks,''
  \emph{IEEE/ACM Transactions on Networking}, vol.~27, no.~2, pp. 662--675,
  2019.

\bibitem{assumption1}
\BIBentryALTinterwordspacing
W.~Ma and D.~Simchi-Levi, ``Tight weight-dependent competitive ratios for
  online edge-weighted bipartite matching and beyond,'' in \emph{Proceedings of
  the 2019 ACM Conference on Economics and Computation}, ser. EC '19.\hskip 1em
  plus 0.5em minus 0.4em\relax New York, NY, USA: Association for Computing
  Machinery, 2019, p. 727–728. [Online]. Available:
  \url{https://doi.org/10.1145/3328526.3329636}
\BIBentrySTDinterwordspacing

\bibitem{posted1}
\BIBentryALTinterwordspacing
R.~Wang, ``Auctions versus posted-price selling,'' \emph{The American Economic
  Review}, vol.~83, no.~4, pp. 838--851, 1993. [Online]. Available:
  \url{http://www.jstor.org/stable/2117581}
\BIBentrySTDinterwordspacing

\bibitem{posted2}
\BIBentryALTinterwordspacing
L.~Einav, C.~Farronato, J.~Levin, and N.~Sundaresan, ``{Auctions versus Posted
  Prices in Online Markets},'' \emph{Journal of Political Economy}, vol. 126,
  no.~1, pp. 178--215, 2018. [Online]. Available:
  \url{https://ideas.repec.org/a/ucp/jpolec/doi10.1086-695529.html}
\BIBentrySTDinterwordspacing

\bibitem{ODE}
\BIBentryALTinterwordspacing
R.~Cooke and V.~Arnold, \emph{Ordinary Differential Equations}, ser. Springer
  Textbook.\hskip 1em plus 0.5em minus 0.4em\relax Springer Berlin Heidelberg,
  1992. [Online]. Available:
  \url{https://books.google.ca/books?id=JUoyqlW7PZgC}
\BIBentrySTDinterwordspacing

\bibitem{competitive}
A.~Borodin and R.~El-Yaniv, \emph{Online computation and competitive
  analysis}.\hskip 1em plus 0.5em minus 0.4em\relax Cambridge University Press,
  2005.

\bibitem{convex-opt}
S.~Boyd and L.~Vandenberghe, \emph{Convex Optimization}.\hskip 1em plus 0.5em
  minus 0.4em\relax USA: Cambridge University Press, 2004.

\bibitem{data}
{GUOSHENG Securities}, ``The way for video platforms to raise the arppu value
  is expected to co-exist with the pressure,''
  \url{http://pg.jrj.com.cn/acc/Res/CN_RES/INDUS/2020/5/17/53fbbf09-3e88-4495-98fc-ba9ae7dbdf81.pdf},
  May 2020.

\bibitem{online-ns-auction}
L.~{Liang}, Y.~{Wu}, G.~{Feng}, X.~{Jian}, and Y.~{Jia}, ``Online auction-based
  resource allocation for service-oriented network slicing,'' \emph{IEEE
  Transactions on Vehicular Technology}, vol.~68, no.~8, pp. 8063--8074, 2019.

\bibitem{5G-archi2}
{NGMN Alliance}, ``5g white paper 2,''
  \url{https://www.ngmn.org/wp-content/uploads/NGMN-5G-White-Paper-2.pdf},
  2020.

\bibitem{auction}
\BIBentryALTinterwordspacing
H.~Ding, J.~Huang, H.~Cao, and Y.~Liu, ``Improving cold music recommendation
  through hierarchical audio alignment,'' in \emph{2013 IEEE 54th Annual
  Symposium on Foundations of Computer Science}.\hskip 1em plus 0.5em minus
  0.4em\relax Los Alamitos, CA, USA: IEEE Computer Society, oct 2016, pp.
  77--82. [Online]. Available:
  \url{https://doi.ieeecomputersociety.org/10.1109/ISM.2016.0023}
\BIBentrySTDinterwordspacing

\end{thebibliography}

\begin{IEEEbiography}
    [{\includegraphics[width=1in,height=1.25in,clip,keepaspectratio]{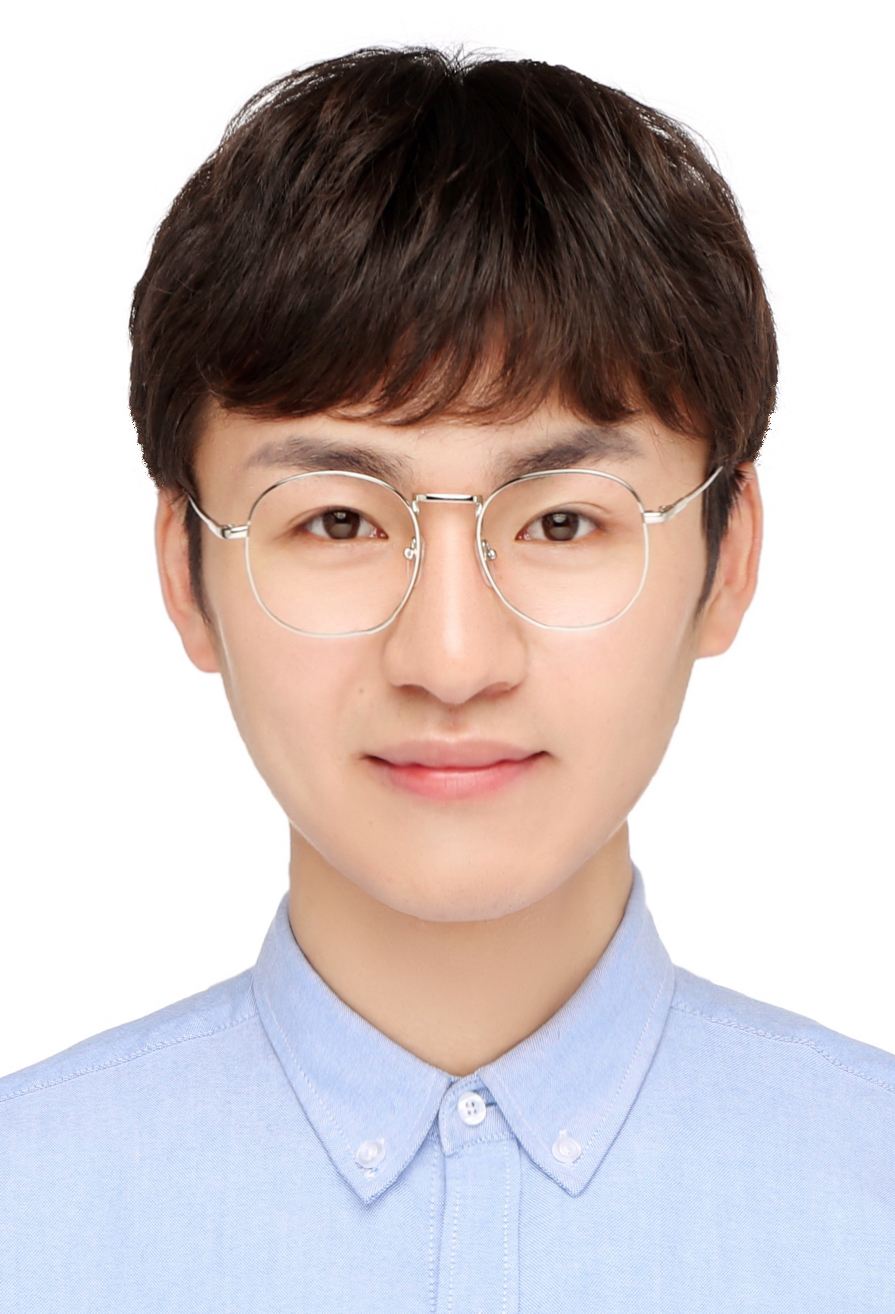}}]{Hailiang Zhao} 
    received the B.S. degree in 2019 from the school of computer science and technology, 
    Wuhan University of Technology, Wuhan, China. He is currently pursuing the Ph.D. degree with the 
    College of Computer Science and Technology, Zhejiang University, Hangzhou, China. He has been a 
    recipient of the Best Student Paper Award of IEEE ICWS 2019. His research interests include edge 
    computing, service computing and machine learning.
\end{IEEEbiography}

\begin{IEEEbiography}
    [{\includegraphics[width=1in,height=1.25in,clip,keepaspectratio]{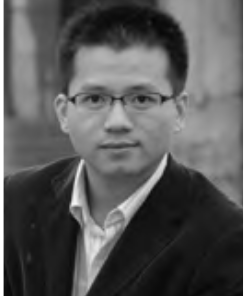}}]{Shuiguang Deng} 
    is currently a full professor at the College of Computer Science and Technology in Zhejiang University, China, 
    where he received a BS and PhD degree both in Computer Science in 2002 and 2007, respectively. He previously 
    worked at the Massachusetts Institute of Technology in 2014 and Stanford University in 2015 as a visiting scholar. 
    His research interests include Edge Computing, Service Computing, Cloud Computing, and Business Process Management. 
    He serves for the journal IEEE Trans. on Services Computing, Knowledge and Information Systems, Computing, and IET 
    Cyber-Physical Systems: Theory \& Applications as an Associate Editor. Up to now, he has published more than 100 
    papers in journals and refereed conferences. In 2018, he was granted the Rising Star Award by IEEE TCSVC. He is 
    a fellow of IET and a senior member of IEEE.
\end{IEEEbiography}

\begin{IEEEbiography}
    [{\includegraphics[width=1in,height=1.25in,clip,keepaspectratio]{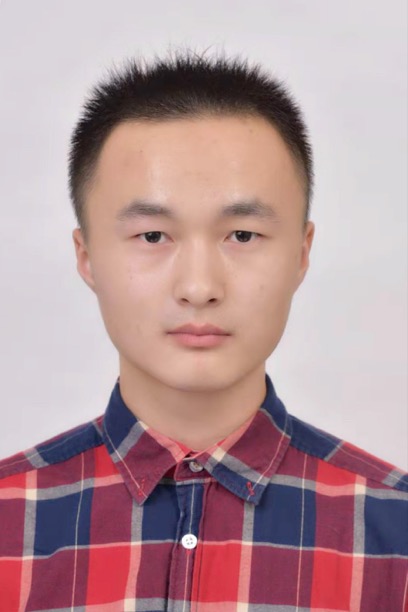}}]{Zijie Liu} 
    received the B.S. degree in 2018 from the school of computer science and technology, 
    Huazhong University of Science an Technology, Wuhan, China. He is now pursuing the master degree 
    with the College of Computer Science and Technology, Zhejiang University, Hangzhou, China. His 
    research interests include edge computing and software engineering.
\end{IEEEbiography}

\begin{IEEEbiography}
    [{\includegraphics[width=1in,height=1.25in,clip,keepaspectratio]{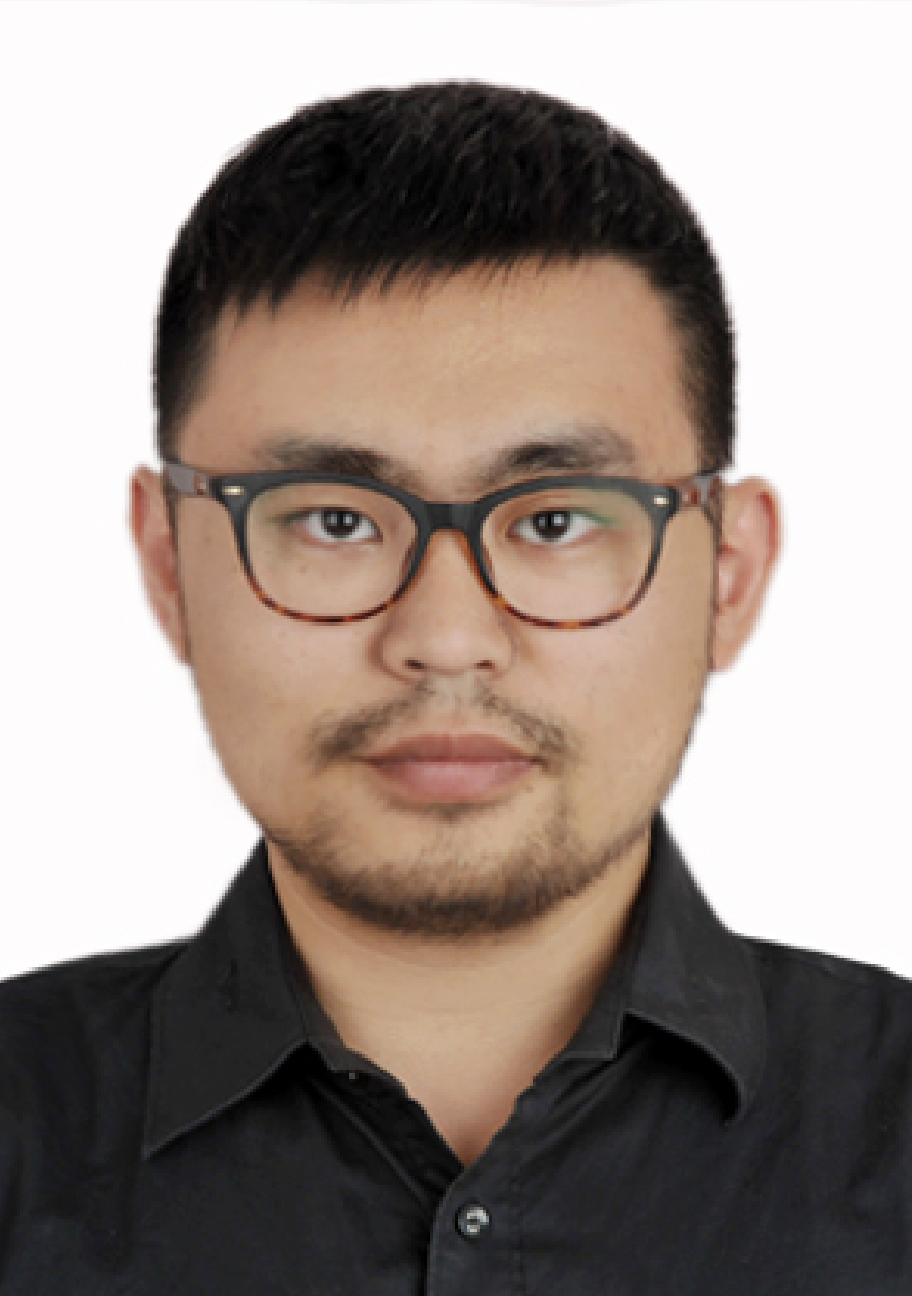}}]{Zhengzhe Xiang}
    received the B.S. and Ph.D. degree of Computer Science and Technology in Zhejiang University, Hangzhou, 
    China. He was previously a visiting student worked at the Karlstad University, Sweden in 2018. He is 
    currently a Lecturer with Zhejiang University City College, Hangzhou, China. His research interests lie in 
    the fields of Service Computing, Cloud Computing, and Edge Computing. 
\end{IEEEbiography}

\begin{IEEEbiography}
    [{\includegraphics[width=1in,height=1.25in,clip,keepaspectratio]{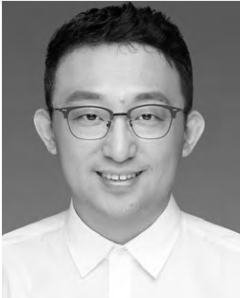}}]{Jianwei Yin} 
    received the Ph.D. degree in computer science from Zhejiang University (ZJU) in 2001. 
    He was a Visiting Scholar with the Georgia Institute of Technology. He is currently a Full Professor 
    with the College of Computer Science, ZJU. Up to now, he has published more than 100 papers in top 
    international journals and conferences. His current research interests include service computing 
    and business process management. He is an Associate Editor of the IEEE Transactions on Services 
    Computing.
\end{IEEEbiography}

\begin{IEEEbiography}[{\includegraphics[width=1in,height=1.25in,clip,keepaspectratio]{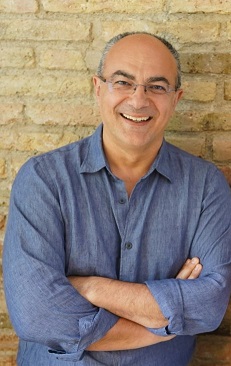}}]{Schahram Dustdar}
    is a Full Professor of Computer Science (Informatics) with a focus on Internet Technologies heading the Decentralized 
    Systems Group at the TU Wien. He is Chairman of the Informatics Section of the Academia Europaea (since December 9, 2016). 
    He is elevated to IEEE Fellow (since January 2016). From 2004-2010 he was Honorary Professor of Information Systems 
    at the Department of Computing Science at the University of Groningen (RuG), The Netherlands.

    From December 2016 until January 2017 he was a Visiting Professor at the University of Sevilla, Spain and from January 
    until June 2017 he was a Visiting Professor at UC Berkeley, USA. He is a member of the IEEE Conference Activities Committee 
    (CAC) (since 2016), of the Section Committee of Informatics of the Academia Europaea (since 2015), a member of the Academia 
    Europaea: The Academy of Europe, Informatics Section (since 2013). He is recipient of the ACM Distinguished Scientist award 
    (2009) and the IBM Faculty Award (2012). He is an Associate Editor of IEEE Transactions on Services Computing, ACM Transactions 
    on the Web, and ACM Transactions on Internet Technology and on the editorial board of IEEE Internet Computing. He is the 
    Editor-in-Chief of Computing (an SCI-ranked journal of Springer).
\end{IEEEbiography}

\begin{IEEEbiography}[{\includegraphics[width=1in,height=1.25in,clip,keepaspectratio]{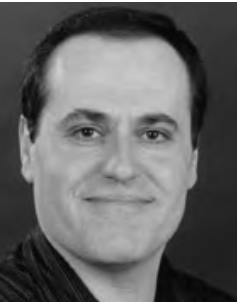}}]{Albert Y. Zomaya}
    is the Chair Professor of High Performance Computing \& Networking in the School of Computer Science, University of 
    Sydney, and he also serves as the Director of the Centre for Decentralized and High Performance Computing. Professor Zomaya 
    published more than 600 scientific papers and articles and is author, co-author or editor of more than 30 books. He is the 
    Editor in Chief of the IEEE Transactions on Sustainable Computing and ACM Computing Surveys and serves as an associate 
    editor for several leading journals. Professor Zomaya served as an Editor in Chief for the IEEE Transactions on Computers 
    (2011-2014). He is a Chartered Engineer, a Fellow of AAAS, IEEE, and IET. Professor Zomaya’s research interests are in the 
    areas of parallel and decentralized computing and complex systems.
  \end{IEEEbiography}

\end{document}